\documentclass[a4paper,12pt]{article}
\usepackage{amsfonts,amsthm,amsmath,amssymb,graphicx,hyperref,youngtab}

\def\a{\alpha}
\def\b{\beta}

\def\m{\mu}

\def\n{\nu}

\def\r{\rho}

\def\s{\sigma}

\def\T{\mathbb{T}}

\newcommand{\cN}{\mathcal N}

\newcommand{\be}{\begin{equation}}
\newcommand{\bea}{\begin{eqnarray}}
\newcommand{\ee}{\end{equation}}
\newcommand{\eea}{\end{eqnarray}}

\newcommand{\nn}{\nonumber}

\def\s{ \sigma}

\begin{document}

\makeatletter
\@addtoreset{equation}{section}
\makeatother
\renewcommand{\theequation}{\thesection.\arabic{equation}}

\rightline{WITS-CTP-123}
\vspace{1.8truecm}

\vspace{15pt}


{\LARGE{  
\centerline{   \bf Subleading corrections to the Double Coset Ansatz} 
\centerline {\bf  preserve integrability} 
}}  

\vskip.5cm 

\thispagestyle{empty} \centerline{
    {\large \bf Robert de Mello Koch\footnote{ {\tt robert@neo.phys.wits.ac.za}}, Stuart Graham\footnote{\tt trautsgraham@gmail.com} 
                and Wandile Mabanga\footnote{\tt Wandile.Mabanga@students.wits.ac.za}} }

\vspace{.4cm}
\centerline{{\it National Institute for Theoretical Physics ,}}
\centerline{{\it Department of Physics and Centre for Theoretical Physics }}
\centerline{{\it University of Witwatersrand, Wits, 2050, } }
\centerline{{\it South Africa } }
\vspace{.4cm}

\vspace{1.4truecm}

\thispagestyle{empty}

\centerline{\bf ABSTRACT}

\vskip.4cm 

In this article we compute the anomalous dimensions for a class of operators, belonging to the $SU(3)$ sector of the theory, 
that have a bare dimension of order $N$.
For these operators the large $N$ limit and the planar limit are distinct and summing only the planar
diagrams will not capture the large $N$ dynamics.
Although the spectrum of anomalous dimensions has been computed for this class of operators, previous studies have
neglected certain terms which were argued to be small.
After dropping these terms diagonalizing the dilatation operator reduces to diagonalizing a set of decoupled oscillators.
In this article we explicitely compute the terms which were neglected previously and show that diagonalizing the dilatation operator 
still reduces to diagonalizing a set of decoupled oscillators.

\setcounter{page}{0}
\setcounter{tocdepth}{2}

\newpage

\tableofcontents

\setcounter{footnote}{0}

\linespread{1.1}
\parskip 4pt

{}~
{}~

\section{Introduction}

Motivated largely by the AdS/CFT correspondence\cite{malda,Gubser:1998bc,Witten:1998qj}, dramatic 
progress in the computation of the spectrum of anomalous dimensions of ${\cal N}=4$ super Yang-Mills theory has been made. 
This has been due, primarily, to the discovery of integrability in the planar limit of the theory\cite{mz,bks,intreview}.
It is interesting to explore the fate of integrability beyond the planar limit, and this is the main motivation for the present study. 
In this article we will compute the anomalous dimensions for a class of operators that have a bare dimension of order $N$.
For these operators, the leading large $N$ limit receives contributions from non-planar diagrams\cite{Balasubramanian:2001nh}. 
For the questions we will study, the large $N$ limit and the planar limit are distinct and summing only the planar
diagrams simply does not capture the large $N$ dynamics.
This is an important point that we will elaborate on below.
As will become clear, methods employing group representation theory have been very effective in solving this 
problem\cite{cjr,dssi,dssii,Kimura:2007wy,bds,BHR1,Bhattacharyya:2008rb,BHR2,countconst,Kimura:2008ac,Kimura:2010tx,pasram,Kimura:2012hp}. 

In this article we focus on operators constructed from the six scalar Higgs fields $\phi_i$ $i=1,2,...,6$ that transform in the 
adjoint of the $U(N)$ gauge group. 
We can form 3 complex combinations
\bea
   X=\phi_1+i\phi_2,\qquad
   Y=\phi_3+i\phi_4,\qquad
   Z=\phi_5+i\phi_6\, .
\eea
Our operators will be constructed using $X$, $Y$ and $Z$.
This is not a consistent truncation of the theory, but it is consistent at one loop (see for example \cite{Hernandez:2004uw}).
The non-zero free field two point functions are
\bea
   \langle Z^i_j(Z^\dagger)^k_l\rangle =\delta^i_l\delta^k_j=\langle Y^i_j(Y^\dagger)^k_l\rangle=\langle X^i_j(X^\dagger)^k_l\rangle\, .
\eea
In writing these two point functions we are not keeping track of spacetime dependence.
Conformal invariance determines the coordinate dependence of the correlator and so we focus on the
the combinatorial problem of summing all possible contractions of the free field propagators. 
The operators we consider are constructed using $n$ $Z$ fields, $m$ $Y$ fields and $p$ $X$ fields.
We are primarily interested in the case that $n,m,p$ all scale as $N$ in the large $N$ limit, but $n\gg m+p$ and
${m\over p}\sim 1$.
Thus, the operators that we consider are a small perturbation of a ${1\over 2}$-BPS operator.
Before we consider this case, we study the much simpler ${1\over 2}$-BPS sector obtained by taking $m=p=0$ with $n\ne 0$. 
Although the dimensions of operators belonging to the ${1\over 2}$-BPS sector are protected so that all anomalous dimensions vanish,
this sector of the theory is rich enough to illustrate much of the structure of the large $N$ limit we study.
The gauge invariant operators we can construct are products of traces of powers of $Z$.
The complete list of observables for the first three values of $n$ is
\bea
n=1:\,\, &&{\rm Tr} Z\, ,\cr
n=2:\,\, &&{\rm Tr} Z^2,\quad {\rm Tr} Z{\rm Tr} Z\, ,\cr
n=3:\,\, &&{\rm Tr} Z^3,\quad {\rm Tr} Z^2{\rm Tr} Z,\quad {\rm Tr} Z{\rm Tr} Z{\rm Tr} Z\, .
\eea
The complete list of observables at any $n$ can clearly be put into one-to-one correspondence with partitions of $n$.
There is a simplification that is only present in the planar limit of the theory: distinct muti-trace structures do not mix.
To illustrate this point, consider the operators
\bea
   O_J\equiv {{\rm Tr} Z^J\over \sqrt{JN^J}}
\eea
which are normalized to have a unit two point function
\bea
   \langle O_J O_J^\dagger\rangle =1\, .
   \label{basic2point}
\eea
To obtain (\ref{basic2point}) we have summed only the planar diagrams.
This is perfectly accurate at large $N$ as long as $J^2\ll N$.
Now consider the two point function between a double trace structure given by $O_{J_1}O_{J_2}$ and the single trace $O_{J_1+J_2}$
\bea
   \langle O_{J_1}O_{J_2}O_{J_1+J_2}^\dagger\rangle = {\sqrt{J_1 J_2 (J_1+J_2)}\over N}\, .
   \label{dynamicalfact}
\eea
If we take $N\to\infty$ holding $J_1$ and $J_2$ fixed, it is clear that the above two point function vanishes.
There is no conservation law forcing this correlator to vanish - its a nontrivial statement about the dynamics.
The two point function in the planar limit, between two operators that have different multitrace structures, vanishes.
Although we have described this only in the ${1\over 2}$-BPS sector and for a specific example, this is a general property of matrix models.
Thus, if we want to compute anomalous dimensions in the planar limit of the theory, we can focus on single trace operators since
these will not mix with operators that have a different trace structure. 
This property of the planar limit is a crucial ingredient in the arguments for the integrability of the planar limit.
Indeed, integrability follows because the planar dilatation operator can be identified with the Hamiltonian of an integrable spin chain. 
A single trace operator containing $K$ fields can be identified with a spin chain state, where the spin chain lives on a lattice that has $K$ sites.
The fields in the single trace operator determine the states of the spins in the lattice.
In this way, there is a bijection between single trace operators and the states of a spin chain.
If we scale $J_1,J_2$ as $N^{2\over 3}$, the right hand side of (\ref{dynamicalfact}) scales as $N^0$ at large $N$ and different trace structures
start to mix.
This mixing sets in even sooner: if we had computed the left hand side of (\ref{dynamicalfact}) exactly, we would find that mixing between different
trace structures is no longer suppressed if $J_1,J_2\gtrsim \sqrt{N}$\cite{Kristjansen:2002bb}.
For the case of interest to us $J_i\sim N$ and there is uncontrolled mixing.
Consequently, the bijection between single trace operators and the states of a spin chain is not useful at all and the link to the
dynamics of a spin chain is lost.
In this article we will describe a way of thinking about the sector of operators that interest us.
In the same way that the spin chain makes the integrability present in the planar limit manifest, our new description shows that the problem
of computing one loop anomalous dimensions is equivalent to solving for the spectrum of a set of decoupled oscillators.

Since the mixing of traces is unconstrained, we need a convenient description that easily allows us to simultaneously talk about all
possible trace structures. 
$Z^i_j$ is an operator acting on $N$ dimensional vector space $V$.
By tensoring $n$ copies of $Z$ we obtain an operator $Z^{\otimes n}\equiv Z\otimes Z\otimes \cdots\otimes Z$ which acts on the space $V^{\otimes n}$. 
$V^{\otimes n}$ admits a natural action of the symmetric group $S_n$ obtained by allowing $\sigma\in S_n$ to permute the factors
of $V$ in $V^{\otimes n}$.
Concretely, for $\sigma\in S_n$ we have
\bea
  (\sigma)^I_J =\delta^{i_1}_{j_{\sigma(1)}}\delta^{i_2}_{j_{\sigma(2)}}\cdots \delta^{i_n}_{j_{\sigma(n)}}\, .
  \label{defsigma}
\eea
Using this action
\bea
  {\rm Tr}(\sigma Z^{\otimes\, n})=\sigma^I_J (Z^{\otimes n})^J_I 
  =Z^{i_1}_{i_{\sigma(1)}}Z^{i_2}_{i_{\sigma(2)}} \cdots Z^{i_n}_{i_{\sigma(n)}}\, .
  \label{singletrace}
\eea 
We can obtain every possible multi-trace structure by choosing the correct permutation $\sigma$. 
For example, consider the simplest non-trivial case $n=2$.
The possible permutations, in cycle notation, are $\sigma =\{(1)(2),(12)\}$ which gives
\bea
   \sigma=(1)(2) && {\rm Tr}(\sigma Z^{\otimes\, 2})=Z^{i_1}_{i_{\sigma(1)}}Z^{i_2}_{i_{\sigma(2)}}
                                                    =Z^{i_1}_{i_1}Z^{i_2}_{i_2}={\rm Tr} Z{\rm Tr} Z\, ,\cr
   \sigma=(12) && {\rm Tr}(\sigma Z^{\otimes\, 2})=Z^{i_1}_{i_{\sigma(1)}}Z^{i_2}_{i_{\sigma(2)}}
                                                    =Z^{i_1}_{i_2}Z^{i_2}_{i_1}={\rm Tr} Z^2\, .
\eea
At $n=2$ each permutation corresponds to a different gauge invariant operator.
This is not generic.
In general permutations in the same conjugacy class determine the same operator.
The set up we have just outlined allows us to trade gauge invariant operators for permutations.
Thus the different multi-trace structures can now be discussed on an equal footing - each corresponds to a permutation.
For the large $N$ limit we consider there is a particularly useful set of gauge invariant operators, known as the Schur polynomials, 
given by \cite{cjr}
\bea
   \chi_R(Z)={1\over n!}\sum_{\sigma\in S_n}\chi_R(\sigma){\rm Tr}(\sigma Z^{\otimes\,\, n})\, .
   \label{Schurdef}
\eea
Notice that the right hand side includes a sum over all possible permutations which implies that the Schur
polynomials are a sum of all possible multi-trace structures.
$\chi_R(\sigma)$ is the character of symmetric group element $\sigma$ in irreducible representation $R$.
The irreducible representations of the symmetric group $S_n$ are labeled by Young diagrams with $n$ boxes.
The set of Young diagrams with $n$ boxes correspond to the partitions of $n$, so that the number of Schur polynomials
matches the number of gauge invariant operators.
The Schur polynomials simply provide an alternative basis to the trace basis.
The fact that the matching of gauge invariant operators matches is more subtle than our discussion above suggests.
Imagine that $N=2$. 
It is easy to check that at $n=3$ there are only two independent gauge invariant operators because (just write the
two sides of this equation out in the basis in which $Z$ is diagonal)
\bea
   {\rm Tr}(Z^3)={1\over 2}\left[ 3{\rm Tr} Z^2 {\rm Tr} Z - {\rm Tr}Z{\rm Tr}Z{\rm Tr}Z\right]\, .
\eea
This is called a trace relation and there will be relations of this type whenever $n>N$ as is the case here.
The Schur polynomials naturally take the trace relations into account, because the Schur polynomial $\chi_R(Z)$ vanishes as soon as
the Young diagram $R$ has more than $N$ rows. Thus, for $N=2$ and $n=3$ there are only two Schur polynomials, given by
$\chi_{\tiny \yng(3)}(Z)$ and $\chi_{\tiny \yng(2,1)}(Z)$. 
Since we are going to take $N\to\infty$ one might expect that the trace relations never apply.
This is the case in the planar limit where the number of fields in our operator is held fixed as we scale $N\to\infty$.
However, for the problems of interest in this article, the number of fields in each operator is also scaled as the limit is taken so that
the number of fields in each operator generically exceeds $N$ and the trace relations must be respected.
Another important property of the Schur polynomials is that they diagonalize the two point function
\bea
   \langle \chi_R(Z)\chi_S(Z^\dagger) \rangle = f_R\delta_{RS}
\eea
where $f_R$ is a product of factors, one for each box in the Young diagram $R$. 
A box in column $j$ and row $i$ has factor $N-i+j$.

The dilatation operator annihilates all operators in the ${1\over 2}$-BPS sector. 
Thus, to obtain a non-trivial anomalous dimension problem we need to move beyond the ${1\over 2}$-BPS sector, by taking 
$p\ne 0$ and/or $m\ne 0$. 

Our discussion above of the ${1\over 2}$-BPS sector generalizes nicely to this more general setting.
Multitrace operators can again be associated to permutations $\sigma\in S_{m+n+p}$
\bea
   {\rm Tr} (\sigma X^{\otimes\, p}\otimes Y^{\otimes\, m}\otimes Z^{\otimes\, n})
   =X^{i_1}_{i_{\sigma(1)}}\cdots X^{i_p}_{i_{\sigma(p)}}Y^{i_{p+1}}_{i_{\sigma(p+1)}}\cdots Y^{i_{p+m}}_{i_{\sigma(p+m)}}\cr
   \times Z^{i_{m+p+1}}_{i_{\sigma(m+p+1)}}\cdots Z^{i_{m+p+n}}_{i_{\sigma(m+p+n)}}\, .
\eea
Permutations that are conjugate, with respect to the $S_p\times S_m\times S_n$ subgroup
\bea
   \gamma\sigma_1\gamma^{-1}=\sigma_2 \qquad \gamma\in S_p\times S_m\times S_n
\eea
give rise to the same operator. 
The Schur polynomials generalize to the restricted Schur polynomials\cite{Balasubramanian:2004nb,Bhattacharyya:2008rb}
\bea
   \chi_{R,(t,s,r)\vec{\mu}\vec{\nu}}={1\over n!m!p!}\sum_{\sigma\in S_{n+m+p}}{\rm Tr}_{(t,s,r)\vec{\mu}\vec{\nu}}(\Gamma^R(\sigma))
   X^{i_1}_{i_{\sigma(1)}}\cdots X^{i_p}_{i_{\sigma(p)}}Y^{i_{p+1}}_{i_{\sigma(p+1)}}\cdots Y^{i_{p+m}}_{i_{\sigma(p+m)}}\cr
   \times Z^{i_{m+p+1}}_{i_{\sigma(m+p+1)}}\cdots Z^{i_{m+p+n}}_{i_{\sigma(m+p+n)}}\, .
\eea
We call ${\rm Tr}_{(t,s,r)\vec{\mu}\vec{\nu}}(\Gamma^R(\sigma))$ the restricted trace of $\Gamma^R(\sigma)$\cite{dssi}.
When computing this trace, we trace over a subspace of the carrier space of $R$.
$R$ is an irreducible representation of $S_{n+m+p}$, that is, it is a Young diagram with $m+n+p$ boxes.
We write $R\vdash m+n+p$.
This subspace we trace over is a carrier space of the subgroup $S_n\times S_m\times S_p$.
It is labeled by three Young diagrams $t\vdash p$, $s\vdash m$ and $r\vdash n$.
$\vec{\mu}$ and $\vec{\nu}$ are degeneracy labels; they are each two dimensional vectors. 
Their two components resolve different copies of the two representations $s\vdash m$ and $t\vdash p$. 
To properly understand the role of the degeneracy labels and what they label, we note that the restricted trace
can be written as
\bea
  {\rm Tr}_{(t,s,r)\vec{\mu}\vec{\nu}}(\cdots )={\rm Tr}_R(P_{(t,s,r)\vec{\mu}\vec{\nu}}\cdots )
\eea
where $P_{(t,s,r)\vec{\mu}\vec{\nu}}$ is an intertwining map.
The degenaracy labels $\vec{\mu}$ and $\vec{\nu}$ play an important role in constructing this intertwining map as we now explain.
The first step in constructing $P_{(t,s,r)\vec{\mu}\vec{\nu}}$ entails constructing a basis for the $(t,s,r)$ irreducible
representation of $S_n\times S_m\times S_p$. 
To do this start from the Young diagram for irreducible representation $R$.
Remove $p$ boxes in any order such that everytime a box is removed what remains is a valid Young diagram and we remove $p_i$
boxes from row $i$. 
Assemble the $p_i$ into a vector $\vec{p}$; this vector will play an important role in what follows.
Now remove $m$ boxes in any order such that everytime a box is removed what remains is a valid Young diagram and we remove $m_i$
boxes from row $i$. 
Assemble the $m_i$ into a vector $\vec{m}$; again, this vector will play an important role in what follows.
The boxes are labeled according to the order in which they are removed so that the first box removed is box 1, the second box removed is box 2, and so on.
In this way we land up with a partly labeled Young diagram $R$.
The unlabeled boxes have the shape $r$ and each partly labeled Young diagram is a distinct subspace of $R$ that carries the irreducible representation
$r$ under the $S_n$ subgroup. 
Now assemble the vectors with first $p$ boxes labeled into an irrep $t$ of $S_p$, resolving multiplicities that arise with $\nu_1$.
In this process, the labels of the next $m$ boxes are simply ignored.
For each state in a given $S_p$ irreducible representation specified by both $t$ and $\nu_1$, one has all possible labelings of the next $m$ boxes.
Assemble these into vectors in an irreducible representation $s$ of $S_m$, resolving multiplicities with $\nu_2$. 
The two multiplicity labels are assembled to produce the vector $\vec{\nu}=(\nu_1,\nu_2)$.
The result of this exercise is a set of vectors labeled with two irreducible representations $t\vdash p$ and $s\vdash m$ each with a multiplicity 
label $\nu_1$ and $\nu_2$, and two state labels, $a,b$, one for each state $|t,\nu_1,a;s,\nu_2,b\rangle$. 
The boxes that are not labeled stand for vectors that belong to a unique irreducible representation $r$ of $S_n$.
Use $c$ to label states in $r$. 
We can make this explicit and write our state as $|t,\nu_1,a;s,\nu_2,b;r,c\rangle$.
This gives a basis for the $(t,s,r)$ irreducible representation of $S_n\times S_m\times S_p$.
Now, the intertwining map is a matrix so that it has both a row label and a column label.
We can use different copies of the $(t,s,r)$ irreducible representation for the rows and columns of the intertwining map.
Consequently
\bea
   P_{(t,s,r)\vec{\mu}\vec{\nu}}=\sum_{a,b,c}|t,\mu_1,a;s,\mu_2,b;r,c\rangle\langle t,\nu_1,a;s,\nu_2,b;r,c|
\eea
Since the $S_m$ and $S_p$ actions commute it is clear that
\bea
   |t,\mu_1,a;s,\mu_2,b;r,c\rangle = |t,\mu_1,a\rangle\otimes |s,\mu_2,b\rangle \otimes |r,c\rangle
\eea
where $\otimes$ is the usual tensor product on a vector space. 
It then also follows that the intertwining maps can be written as a tensor product
\bea
  P_{(t,s,r)\vec{\mu}\vec{\nu}}&=&\sum_{a}|t,\mu_1,a\rangle\langle t,\nu_1,a|\otimes \sum_{b}|s,\mu_2,b\rangle\langle s,\nu_2,b|\otimes
                                \sum_{c}|r,c\rangle\langle r,c|\cr
            &\equiv& p_{t\mu_1\nu_1}\otimes p_{s\mu_2\nu_2}\otimes {\bf 1}_r
\eea
The last factor in this product is always a genuine projector.

The restricted Schur polynomials share many of the nice properties that make the Schur polynomials so useful.
In particular, the restricted Schur polynomials respect the trace relations and the two point function of the restricted Schur 
polynomials\cite{Bhattacharyya:2008rb}
\bea
  \langle \chi_{R,(t,s,r)\vec{\mu}\vec{\nu}} \chi^\dagger_{T,(y,x,w)\vec{\b}\vec{\a}} \rangle
    ={f_R {\rm hooks}_R \over {\rm hooks}_r {\rm hooks}_s {\rm hooks}_t}\delta_{RT}\delta_{rw}\delta_{sx}\delta_{ty}
     \delta_{\vec{\m}\vec{\b}}\delta_{\vec{\n}\vec{\a}}
\eea
again diagonalize the free field two point function.
The number ${\rm hooks}_R$ is a product of the hook lengths in Young diagram $R$.
We will often find it convenient to work with operators $\hat{O}_{R,(t,s,r)\vec{\mu}\vec{\nu}}$ normalized to have a unit two point function.
These operators are related to the restricted Schur polynomials $\chi_{R,(t,s,r)\vec{\mu}\vec{\nu}}$ as
\bea
 \hat{O}_{R,(t,s,r)\vec{\mu}\vec{\nu}}=\sqrt{ {\rm hooks}_r {\rm hooks}_s {\rm hooks}_t \over f_R {\rm hooks}_R}
                                       \chi_{R,(t,s,r)\vec{\mu}\vec{\nu}}\, .
\eea

The key difficulty with working with the restricted Schur polynomials, is in constructing and working with the intertwining maps 
$P_{(t,s,r)\vec{\mu}\vec{\nu}}$.
Convenient methods to accomplish this have been developed for two rows in \cite{Carlson:2011hy} and in general in \cite{Koch:2011hb}.
Using these methods, the one loop dilatation operator has been diagonalized in the $su(2)$ sector 
(obtained by setting $p=0$)\cite{Carlson:2011hy,Koch:2011hb,gs,deMelloKoch:2012ck}.
In this sector, the one loop dilatation operator reduces to a set of decoupled oscillators, which is an integrable system.
These results provided perfect confirmation of earlier numerical studies\cite{Koch:2010gp,DeComarmond:2010ie}.
At two loops the system remains integrable in the $su(2)$ sector\cite{deMelloKoch:2012sv}.
The one loop results were generalized to $p\ne 0$ in \cite{Koch:2011jk}, but the interactions between the $X$ and $Y$ fields were argued to 
be subleading and were neglected. 
The subleading terms are of order ${m\over n}$ relative to the leading terms\cite{Koch:2011jk}. 
It is precisely these terms that we will evaluate in this paper.

When interactions between the $X$ and $Y$ fields are neglected, the vectors $\vec{p}$ and $\vec{m}$ defined above are conserved\cite{Koch:2011hb}.
The dilatation operator only mixes operators that have the same $\vec{p}$ and $\vec{m}$ values.
This is not at all surprising - integrable systems are always accompanied with higher conserved quantities.
What makes the interaction between the $X$ and $Y$ fields so interesting is that they spoil the conservation of $\vec{p}$ and $\vec{m}$.
This can mean one of two things: either, integrability does not persist beyond the $su(2)$ sector and this large $N$ but non-planar
limit is not integrable, or the dynamics remains integrable but the conservation of $\vec{p}$ and $\vec{m}$ is not one of the
conservation laws of this (extended) integrable system.
Our results are unambiguous - the second case is realized and the one-loop dilatation operator continues to be integrable when
extended to act on operators built using all three complex scalars. 
Indeed, we are able to identify the new terms we have evaluated with elements of the Lie algebra of a unitary group.
Diagonalizing the complete dilatation operator then reduces to a solved problem in representation theory.

This article is organized as follows: 
Section 2 evaluates the action of the one loop dilatation operator on the fields we study in this article.
The description of the operators (the Gauss operators) that diagonalize the terms in the dilatation operator that mix $X$ and $Z$ fields
or $Y$ and $Z$ fields is reviewed in section 3. 
In section 4 we compute the action of the terms in the dilatation operator that mix $X$ and $Y$, on the Gauss operators.
Section 5 is used to argue that the dilatation operator can be written as an element of the Lie algebra of a unitary group.
Our conclusions are given in section 6.

\section{Dilatation Operator}

The one loop dilatation operator in the sector we consider, is given by\cite{mz}
\bea
   D=-g_{YM}^2{\rm Tr}\left(
                     \left[Y,Z\right]\left[{d\over dY},{d\over dZ}\right]
                 +   \left[X,Z\right]\left[{d\over dX},{d\over dZ}\right]
                 +   \left[Y,X\right]\left[{d\over dY},{d\over dX}\right]
                      \right)\, .
\eea
To be completely explicit, the index structure is
\bea
   {\rm Tr}\left( \left[Y,X\right]\left[{d\over dY},{d\over dX}\right]\right)  =
   (Y^i_l X^l_j - X^i_l Y^l_j)\left( {d\over dY^k_j}{d\over dX^i_k} - {d\over dX^k_j}{d\over dY^i_k}\right)\, .
\eea
Our first task is to consider the action of $D$ on restricted Schur polynomials.
In what follows we will often need the identity \cite{rajmike}
\bea
   {\rm Tr}(\sigma Y^{\otimes m}\otimes Z^{\otimes n})=\sum_{T,t,u,\vec{\nu}}{d_T n! m!\over d_t d_u (n+m)!}
      \chi_{T,(t,u)\vec{\nu}^*}(\sigma^{-1})
      \chi_{T,(t,u)\vec{\nu}}(Z,Y)
\eea
where if $\vec{\nu}=(\nu_1,\nu_2)$ then $\vec{\nu}^*=(\nu_2,\nu_1)$.
With a suitable choice of $\sigma$, the right hand side above gives any desired multitrace operator.
Thus, the above equation expresses an arbitrary multitrace operator as a linear combination of restricted Schur polynomials.
The sum above runs over all Young diagrams $T\vdash m+n$, $t\vdash n$ and $u\vdash m$ as well as over the multplicity labels $\vec{\nu}$.
$d_T$ denotes the dimension of the irreducible representation $T$ of $S_{n+m}$.
Similarly, $d_t$ denotes the dimension of irreducible representation $t$ of $S_{n}$ and $d_u$ the dimension of irreducible representation $u$ of $S_{m}$.
Finally, $\chi_{T,(t,u)\vec{\nu}^*}(\sigma^{-1})$ is the restricted character obtained by tracing $\Gamma_R(\sigma^{-1})$ over the
$(t,u)$ subspace, i.e. $\chi_{T,(t,u)\vec{\nu}^*}(\sigma^{-1})={\rm Tr}_{(t,u)\vec{\nu}^*}(\Gamma_T(\sigma^{-1}))$.
The multiplicity index $\vec{\nu}^*=(\nu_2,\nu_1)$ tells us to trace the row index over the $\nu_2$ copy of $(r,s)$ and the column index over the
$\nu_1$ copy.
We will consider in detail the subleading term which mixes $Y$ and $X$. 
The remaining terms can be evaluated in an identical way.
A straight forward computation gives
{\footnotesize
\bea
[Y,X]^i_j\left({d\over dY^k_j}{d\over dX^i_k} - {d\over dX^k_j}{d\over dY^i_k}\right)\chi_{R,(t,s,r)\vec{\mu}\vec{\nu}}\cr
   =[Y,X]^i_j\left({d\over dY^k_j}{d\over dX^i_k} - {d\over dX^k_j}{d\over dY^i_k}\right)
   {1\over n!m!p!}\sum_{\sigma\in S_{n+m+p}}{\rm Tr}_{(t,s,r)\vec{\mu}\vec{\nu}}(\Gamma^R(\sigma))
   X^{i_1}_{i_{\sigma(1)}}\cdots X^{i_p}_{i_{\sigma(p)}}Y^{i_{p+1}}_{i_{\sigma(p+1)}}\cdots Y^{i_{p+m}}_{i_{\sigma(p+m)}}\cr
   \times Z^{i_{m+p+1}}_{i_{\sigma(m+p+1)}}\cdots Z^{i_{m+p+n}}_{i_{\sigma(m+p+n)}}\cr
   \cr
   ={mp\over n!m!p!}\sum_{\sigma\in S_{n+m+p}}{\rm Tr}_{(t,s,r)\vec{\mu}\vec{\nu}}(\Gamma^R(\sigma))
   (\delta^{i_{p+1}}_{i_{\sigma(1)}}[Y,X]^{i_1}_{i_{\s(p+1)}}-\delta^{i_{1}}_{i_{\sigma(p+1)}}[Y,X]^{i_{p+1}}_{i_{\s(1)}})\cr
   \times
   X^{i_2}_{i_{\sigma(2)}}\cdots X^{i_p}_{i_{\sigma(p)}}Y^{i_{p+2}}_{i_{\sigma(p+2)}}\cdots Y^{i_{p+m}}_{i_{\sigma(p+m)}}
   Z^{i_{m+p+1}}_{i_{\sigma(m+p+1)}}\cdots Z^{i_{m+p+n}}_{i_{\sigma(m+p+n)}}\cr
   \cr
   ={mp\over n!m!p!}\sum_{\sigma\in S_{n+m+p}}{\rm Tr}_{(t,s,r)\vec{\mu}\vec{\nu}}(\Gamma^R([(1,p+1),\sigma]))
   \delta^{i_{1}}_{i_{\sigma(1)}}
   X^{i_2}_{i_{\sigma(2)}}\cdots X^{i_p}_{i_{\sigma(p)}}\cr
   \times [Y,X]^{i_{p+1}}_{i_{\s(p+1)}}
   Y^{i_{p+2}}_{i_{\sigma(p+2)}}\cdots Y^{i_{p+m}}_{i_{\sigma(p+m)}}
   Z^{i_{m+p+1}}_{i_{\sigma(m+p+1)}}\cdots Z^{i_{m+p+n}}_{i_{\sigma(m+p+n)}}\nonumber
\eea
}
The delta function in the summand will restrict the sum over $S_{n+m+p}$ to a sum over the $S_{n+m+p-1}$ subgroup.
The $S_{n+m+p-1}$ subgroup is obtained by retaining those elements that hold $i_1$ inert, i.e. $\sigma(1)=1$.
To see how this happens, introduce the notation $\r_i=\s(i,1)$ and rewrite the above sum as a sum over $S_{n+m+p-1}$ and its cosets.
The result is
{\footnotesize
\bea
[Y,X]^i_j\left({d\over dY^k_j}{d\over dX^i_k} - {d\over dX^k_j}{d\over dY^i_k}\right)\chi_{R,(t,s,r)\vec{\mu}\vec{\nu}}\cr
   ={mp\over n!m!p!}\sum_{\sigma\in S_{n+m+p-1}}\sum_{i=1}^{n+m+p}{\rm Tr}_{(t,s,r)\vec{\mu}\vec{\nu}}(\Gamma^R([(1,p+1),\r_i ]))
   \delta^{i_{1}}_{i_{\r_i(1)}}
   X^{i_2}_{i_{\r_i(2)}}\cdots X^{i_p}_{i_{\r_i(p)}}\cr
   \times [Y,X]^{i_{p+1}}_{i_{\r_i(p+1)}}
   Y^{i_{p+2}}_{i_{\r_i(p+2)}}\cdots Y^{i_{p+m}}_{i_{\r_i(p+m)}}
   Z^{i_{m+p+1}}_{i_{\r_i(m+p+1)}}\cdots Z^{i_{m+p+n}}_{i_{\r_i(m+p+n)}}\cr
   \cr
   ={mp\over n!m!p!}\sum_{\sigma\in S_{n+m+p-1}}{\rm Tr}_{(t,s,r)\vec{\mu}\vec{\nu}}(\Gamma^R([(1,p+1),\{ N+\sum_{i=2}^{n+m+p}(i,1)\} ]))\cr
   \times {\rm Tr}_{V^{\otimes\, n+m+p-1}}(\s\cdot X^{\otimes\, p-1}\otimes [Y,X]\otimes Y^{\otimes\, m-1}\otimes Z^{\otimes\, n})\cr
   \cr
   ={mp\over n!m!p!}\sum_{R'} c_{RR'}\sum_{\sigma\in S_{n+m+p-1}}
   {\rm Tr}_{(t,s,r)\vec{\mu}\vec{\nu}}(([\Gamma^R((1,p+1)),\Gamma^{R'}(\s)])\cr
   \times {\rm Tr}_{V^{\otimes\, n+m+p-1}}(\s\cdot X^{\otimes\, p-1}\otimes [Y,X]\otimes Y^{\otimes\, m-1}\otimes Z^{\otimes\, n})\cr
\nonumber
\eea
}
The sum over $R'$ runs over all irreducible representations $R'$ of the $S_{n+m+p-1}$ subgroup that can be subduced from the irreducible
representation $R$ of the $S_{n+m+p}$ subgroup.
As a Young diagram $R'$ is obtained from $R$ by dropping a single box.
A prime on a letter denoting a Young diagram will always indicate that we drop a box.
To obtain the last line above, use the fact that $N+\sum_{i=2}^{n+m+p}(i,1)$ when acting on any state within the subspace $R'$ subduced by $R$ 
gives $c_{RR'}$. 
This is proved by noting that $\sum_{i=2}^{n+m+p}(i,1)$ is a Jucys-Murphy element; see \cite{dssi} for the details.
{\footnotesize
\bea
[Y,X]^i_j\left({d\over dY^k_j}{d\over dX^i_k} - {d\over dX^k_j}{d\over dY^i_k}\right)\chi_{R,(t,s,r)\vec{\mu}\vec{\nu}}\cr
   ={mp\over n!m!p!}\sum_{R'} c_{RR'}\sum_{\sigma\in S_{n+m+p-1}}
   {\rm Tr}_{(t,s,r)\vec{\mu}\vec{\nu}}(([\Gamma^R((1,p+1)),\Gamma^{R'}(\s)])\cr
   \times {\rm Tr}_{V^{\otimes\, n+m+p-1}}([(1,p+1),\s]\cdot X^{\otimes\, p}\otimes Y^{\otimes\, m}\otimes Z^{\otimes\, n})\cr
   ={mp\over n!m!p!}\sum_{T,(y,x,w)\vec{\a}\vec{\b}}{d_T n! m! p!\over d_w d_x d_y (n+m+p)!}\sum_{R'} c_{RR'}\sum_{\sigma\in S_{n+m+p-1}}
   {\rm Tr}_{(t,s,r)\vec{\mu}\vec{\nu}}(([\Gamma^R((1,p+1)),\Gamma^{R'}(\s)])\cr
   {\rm Tr}_{(y,x,w)\vec{\a}\vec{\b}}(\Gamma_T([(1,p+1),\s]))\chi_{T,(y,x,w)\vec{\b}\vec{\a}}(X,Y,Z)\cr
   \cr
   =\sum_{R'} c_{RR'}\sum_{T,(y,x,w)\vec{\a}\vec{\b}} {d_T m p\over d_w d_x d_y (n+m+p) d_{R'}}\cr
    {\rm Tr}_{R\oplus T}([P_{R,(t,s,r)\vec{\mu}\vec{\nu}},\Gamma^R(1,p+1)]I_{R'T'}
                         [P_{T,(y,x,w)\vec{\a}\vec{\b}},  \Gamma^T(1,p+1)]I_{T'R'})\chi_{T,(y,x,w)\vec{\b}\vec{\a}}(X,Y,Z)\, .
\nonumber
\eea
}

To get to the last line sum over $S_{n+m+p-1}$ using the fundamental orthogonality relation.
Writing this result in terms of normalized operators we have
{\footnotesize
\bea
   [Y,X]^i_j\left({d\over dY^k_j}{d\over dX^i_k} - {d\over dX^k_j}{d\over dY^i_k}\right)\hat{O}_{R,(t,s,r)\vec{\mu}\vec{\nu}}\cr
   =\sum_{R'} c_{RR'}\sum_{T,(y,x,w)\vec{\a}\vec{\b}}
    {d_T mp\over d_w d_x d_y (n+m+p)d_{R'}}
    \sqrt{f_T {\rm hooks}_T{\rm hooks}_r{\rm hooks}_s{\rm hooks}_t\over f_R {\rm hooks}_R{\rm hooks}_w{\rm hooks}_x{\rm hooks}_y}\cr
    {\rm Tr}_{R\oplus T}([P_{R,(t,s,r)\vec{\mu}\vec{\nu}},\Gamma^R(1,p+1)]I_{R'T'}
                         [P_{T,(y,x,w)\vec{\a}\vec{\b}},  \Gamma^T(1,p+1)]I_{T'R'})
                         \hat{O}_{T,(y,x,w)\vec{\b}\vec{\a}}\, .\cr
\eea
}
Using identical methods it is straight forward to find
{\footnotesize
\bea
   [Y,Z]^i_j\left({d\over dY^k_j}{d\over dZ^i_k} - {d\over dZ^k_j}{d\over dY^i_k}\right)\hat{O}_{R,(t,s,r)\vec{\mu}\vec{\nu}}\cr
   =\sum_{R'} c_{RR'}\sum_{T,(y,x,w)\vec{\a}\vec{\b}}
    {d_T mn\over d_w d_x d_y (n+m+p)d_{R'}}
    \sqrt{f_T {\rm hooks}_T{\rm hooks}_r{\rm hooks}_s{\rm hooks}_t\over f_R {\rm hooks}_R{\rm hooks}_w{\rm hooks}_x{\rm hooks}_y}\cr
        {\rm Tr}_{R\oplus T}([\Gamma^R(1,p+1)P_{R,(t,s,r)\vec{\mu}\vec{\nu}}\Gamma^R(1,p+1),\Gamma^R(1,p+m+1)]I_{R'T'}\cr
                  \times [\Gamma^T(1,p+1)P_{T,(y,x,w)\vec{\a}\vec{\b}}\Gamma^T(1,p+1),  \Gamma^T(1,p+m+1)]I_{T'R'})
                         \hat{O}_{T,(y,x,w)\vec{\b}\vec{\a}}\, ,\cr
\eea
}
{\footnotesize
\bea
   [X,Z]^i_j\left({d\over dX^k_j}{d\over dZ^i_k} - {d\over dZ^k_j}{d\over dX^i_k}\right)\hat{O}_{R,(t,s,r)\vec{\mu}\vec{\nu}}\cr
   =\sum_{R'} c_{RR'}\sum_{T,(y,x,w)\vec{\a}\vec{\b}}
    {d_T pn\over d_w d_x d_y (n+m+p)d_{R'}}
    \sqrt{f_T {\rm hooks}_T{\rm hooks}_r{\rm hooks}_s{\rm hooks}_t\over f_R {\rm hooks}_R{\rm hooks}_w{\rm hooks}_x{\rm hooks}_y}\cr
    {\rm Tr}_{R\oplus T}([P_{R,(t,s,r)\vec{\mu}\vec{\nu}},\Gamma^R(1,p+m+1)]I_{R'T'}
                         [P_{T,(y,x,w)\vec{\a}\vec{\b}},  \Gamma^T(1,p+m+1)]I_{T'R'})
                         \hat{O}_{T,(y,x,w)\vec{\b}\vec{\a}}\, .\cr
\eea
}

The next step in the evaluation of the action of the dilatation operator entails computing the traces over
$R\oplus T$ that have appeared in our results above. 
Our results for the action of the one loop dilatation operator given above are exact.
From this point on we assume the displaced corners approximation so that our answers for the traces are only
valid in the large $N$ limit.
The reader wanting to follow all of the details in this section should consult \cite{Koch:2011hb} for background.
For the term in the one loop dilatation operator that mixes $X$ and $Y$ the trace that needs to be computed is
\bea
\T={\rm Tr}_{R\oplus T}([P_{R,(t,s,r)\vec{\mu}\vec{\nu}},\Gamma^R(1,p+1)]I_{R'T'}
                         [P_{T,(y,x,w)\vec{\a}\vec{\b}},  \Gamma^T(1,p+1)]I_{T'R'})\, .
\eea
To ease the notation we will use the following shorthand
\bea
  P_{T,(y,x,w)\vec{\a}\vec{\b}}\equiv p_{y}\otimes p_{x}\otimes {\bf 1}_w\, .
\eea
Consider the case that $R'$ is obtained from $R$ by dropping a box in row $i$ and that $T'$ is obtained from $T$ by dropping a box from row $j$.
The intertwiner is only non-zero if $T'=R'$.
In this case the intertwiners are
\bea
   I_{R'T'}=E^{(1)}_{ij},\qquad I_{T'R'}=E^{(1)}_{ji}\, .
\eea
Since the trace $\T$ is a product of two commutators, when we expand things out we get a total of four terms.
Since both the swaps $\Gamma^R(1,p+1)$ and $\Gamma^T(1,p+1)$ have a trivial action on the $Z$ indices, we know that
the result will be proportional to $\delta_{rw}$ and that the trace over the $Z$ indices produce a factor $d_r$.
Thus, after tracing over the $Z$ indices we have 
\bea
&&\T= \left( {\rm Tr}(p_t\otimes p_s  \Gamma^R(1,p+1) E^{(1)}_{ij} p_y\otimes p_x \Gamma^T(1,p+1) E^{(1)}_{ji})\right.\cr
&&         -{\rm Tr}(p_t\otimes p_s  \Gamma^R(1,p+1) E^{(1)}_{ij} \Gamma^T(1,p+1) p_y\otimes p_x E^{(1)}_{ji})\cr
&&         -{\rm Tr}(\Gamma^R(1,p+1) p_t\otimes p_s  E^{(1)}_{ij} p_y\otimes p_x \Gamma^T(1,p+1) E^{(1)}_{ji})\cr
&&\left.
           +{\rm Tr}(\Gamma^R(1,p+1) p_t\otimes p_s  E^{(1)}_{ij} \Gamma^T(1,p+1) p_y\otimes p_x E^{(1)}_{ji})\right)\delta_{rw}d_r\, .
\eea
Allow the swaps to act on the intertwiners
\bea
   (1,p+1)E^{(1)}_{ij}=E^{(1)}_{lj}E^{(p+1)}_{il},\qquad    (1,p+1)E^{(1)}_{ji}=E^{(1)}_{li}E^{(p+1)}_{jl}
\eea
to obtain
\bea
&&\T
=\Big(
  \langle\vec{p},t,\nu_1;a   |E^{(1)}_{lj}  |\vec{p}',y,\alpha_1;b\rangle
  \langle\vec{p}',y,\beta_1;b|E^{(1)}_{ki}  |\vec{p},t,\mu_1;a\rangle\cr
&&\qquad\times
  \langle\vec{m},s,\nu_2;c   |E^{(p+1)}_{il}|\vec{m}',x,\alpha_2;d\rangle
  \langle\vec{m}',x,\beta_2;d|E^{(p+1)}_{jk}|\vec{m},s,\mu_2;c\rangle\cr
&&-\delta_{\vec{p}\vec{p}'}\delta_{yt}\delta_{\nu_1\alpha_1}\delta_{\vec{m}\vec{m}'}\delta_{sx}\delta_{\beta_2\mu_2}
  \langle\vec{p}',y,\beta_1;b|E^{(1)}_{ji}  |\vec{p},t,\mu_1;a\rangle\cr
&&\qquad\times
  \langle\vec{m},s,\nu_2;c   |E^{(p+1)}_{ij}|\vec{m}',x,\alpha_2;d\rangle\cr
&&-\delta_{\vec{p}\vec{p}'}\delta_{yt}\delta_{\mu_1\beta_1}\delta_{\vec{m}\vec{m}'}\delta_{sx}\delta_{\alpha_2\nu_2}
  \langle\vec{p},t,\nu_1;a  |E^{(1)}_{ij} |\vec{p}',y,\alpha_1;b\rangle\cr
&&\qquad\times
  \langle\vec{m}',x,\beta_2;d|E^{(p+1)}_{ji}|\vec{m},s,\mu_2;c\rangle\cr
&&+\langle\vec{p},t,\nu_1;a   |E^{(1)}_{il}  |\vec{p}',y,\alpha_1;b\rangle
  \langle\vec{p}',y,\beta_1;b |E^{(1)}_{jk}|\vec{p},t,\mu_1;a\rangle\cr
&&\qquad\times
  \langle\vec{m},s,\nu_2;c    |E^{(p+1)}_{lj}|\vec{m}',x,\alpha_2;d\rangle
  \langle\vec{m}',x,\beta_2;d |E^{(p+1)}_{ki}|\vec{m},s,\mu_2;c\rangle\Big)\delta_{rw}d_r\, .
\label{elementstoevaluate}
\eea

In a similar way we obtain
\bea
&&{\rm Tr}_{R\oplus T}([\Gamma^R(1,p+1)P_{R,(t,s,r)\vec{\mu}\vec{\nu}}\Gamma^R(1,p+1),\Gamma^R(1,p+m+1)]I_{R'T'}\cr
&&\times [\Gamma^T(1,p+1)P_{T,(y,x,w)\vec{\a}\vec{\b}}\Gamma^T(1,p+1),  \Gamma^T(1,p+m+1)]I_{T'R'})\cr
&&=\delta_{ty}d_t \delta_{\vec{p}\vec{p}'}\delta_{\nu_1\alpha_1}\delta_{\beta_1\mu_1}d_{r_i'}\delta_{r_i'w_j'}\delta_{\vec{m}\vec{m}'}
   \Big[ 
\langle \vec{m}',x,\beta_2;d|E_{ii}^{(p+1)}|\vec{m},s,\mu_2;c\rangle
\langle \vec{m},s\nu_2;c|E_{jj}^{(p+1)}|\vec{m}',x,\alpha_2;d\rangle\cr
&&+\langle \vec{m}',x,\beta_2;d|E_{jj}^{(p+1)}|\vec{m},s,\mu_2;c\rangle
   \langle \vec{m},s\nu_2;c|E_{ii}^{(p+1)}|\vec{m}',x,\alpha_2;d\rangle\Big]\cr
&&-\delta_{ij}\delta_{ty}d_t\delta_{\vec{p}\vec{p}'}\delta_{\nu_1\alpha_1}\delta_{\beta_1,\mu_1}d_{r'_i}\delta_{rw}\delta_{sx}\delta_{\vec{m}\vec{m}'}
         \Big[\delta_{\nu_2\alpha_2}\langle \vec{m},x,\beta_2;c|E_{jj}^{(p+1)}|\vec{m},s,\mu_2;c\rangle\cr
&&+            \delta_{\beta_2\mu_2}\langle \vec{m},s\nu_2;c|E_{ii}^{(p+1)}|\vec{m},x,\alpha_2;c\rangle\Big]
\eea
relevant for the term in the one loop dilatation operator that mixes $Z$ and $Y$ and
\bea
&&{\rm Tr}_{R\oplus T}([P_{R,(t,s,r)\vec{\mu}\vec{\nu}},\Gamma^R(1,p+m+1)]I_{R'T'}
                         [P_{T,(y,x,w)\vec{\a}\vec{\b}},  \Gamma^T(1,p+m+1)]I_{T'R'})\cr
&&=\delta_{sx}d_s \delta_{\vec{p}\vec{p}'}\delta_{\nu_2\alpha_2}\delta_{\beta_2\mu_2}d_{r_i'}\delta_{r_i'w_j'}\delta_{\vec{m}\vec{m}'}
   \Big[ 
\langle \vec{p}',y,\beta_1;d|E_{ii}^{(1)}|\vec{p},t,\mu_1;c\rangle
\langle \vec{p},t,\nu_1;c|E_{jj}^{(1)}|\vec{p}',y,\alpha_1;d\rangle\cr
&&+\langle \vec{p}',y,\beta_1;d|E_{jj}^{(1)}|\vec{p},t,\mu_1;c\rangle
   \langle \vec{p},t,\nu_1;c|E_{ii}^{(1)}|\vec{p}',y,\alpha_1;d\rangle\Big]\cr
&&-\delta_{ij}\delta_{ty}d_s\delta_{\vec{p}\vec{p}'}\delta_{\nu_2\alpha_2}\delta_{\beta_2\mu_2}d_{r'_i}\delta_{rw}\delta_{sx}\delta_{\vec{m}\vec{m}'}
         \Big[\delta_{\nu_1\alpha_1}\langle \vec{p},y,\beta_1;c|E_{jj}^{(1)}|\vec{p},t,\mu_1;c\rangle\cr
&&+            \delta_{\beta_1\mu_1}\langle \vec{p},t,\nu_1;c|E_{ii}^{(1)}|\vec{p},y,\alpha_1;c\rangle\Big]
\eea
which is relevant for the term in the one loop dilatation operator that mixes $X$ and $Z$.

This completes our discussion of the action of the one loop dilatation operator.

\section{Gauss Operators}

The problem of diagonalizing the terms in the dilatation operator that mix the $X$ and $Z$ fields and the terms that mix the $Y$ and $Z$ fields
has been solved\cite{Carlson:2011hy,Koch:2011hb,gs,deMelloKoch:2012ck}.
The operators that have a good scaling dimension are the Gauss operators. 
Our ultimate goal is to write the action of the terms in the dilatation operator that mix $X$ and $Y$ fields, on the Gauss operators, which
amounts to a change of basis from restricted Schur polynomials to Gauss operators.
Towards this end we describe how to construct Gauss operators for operators built from three complex scalar fields and develop the tools
we will need to change basis.
The results of this section are a simple generalization of \cite{deMelloKoch:2012ck}.

Natural hints for the construction of the Gauss operators come from the AdS/CFT correspondence.
Indeed, the correspondence implies an equivalence between quantum states in the quantum gravity and operators in the $\cN=4$ super-Yang-Mills theory.
In particular, the restricted Schur polynomials $\chi_{R,(t,s,r),\vec{\mu}\vec{\nu}}(X,Y,Z)$ are dual to multiple giant graviton
systems \cite{mst,myers,hash}
consisting of large branes in the $AdS_5$ space when $R$ has order one rows each of length order $N$, 
or to systems consisting of large branes in the $S^5$ space when $R$ has  order one columns each of length order $N$.
A giant graviton has a compact world volume so that the Gauss Law forces the total charge on the giant's world volume to vanish. 
Since the string end points are charged, this gives a constraint on the possible open string configurations that are allowed: 
the number of strings leaving the giant must equal the number of strings arriving at the giant.
The matrices $X$ and $Y$ generate two species of 1-bit strings \cite{Giles:1977mpa,Berenstein:2002jq,Balasubramanian:2002sa,Vaman:2002ka}. 
Each row of $R$ corresponds to a giant graviton.
Each open string configuration corresponds to a pair of graphs - one for each open string species.
We will refer to these as the $X$ graph and the $Y$ graph.
The vertices of the graph represent the branes and the directed links represent the (oriented) strings. 
Motivated by \cite{countFeyn} a useful combinatoric description of these graphs is to divide each string into two halves and label each half.
Using the orientation of the string, label the outgoing ends with numbers $\{1,\cdots ,p\}$ for the $X$ graph or $\{1, \cdots , m\}$ for the $Y$ graph 
and the ingoing ends with these same numbers. 
A permutation $\sigma \in S_p\times S_m$ is then determined by how the halves are joined.
We will often decompose $\sigma =\sigma^X\circ \sigma^Y$ with $\sigma^X\in S_p$ and $\sigma^Y\in S_m$.
Given a permutation, we can reconstruct the graphs.
A graph is not associated to a unique permutation because the strings leaving the $i$'th vertex are indistinguishable, 
and the strings arriving at the $i$'th vertex are indistinguishable. 
\begin{figure}[h]
\begin{center}
\resizebox{!}{9cm}{\includegraphics{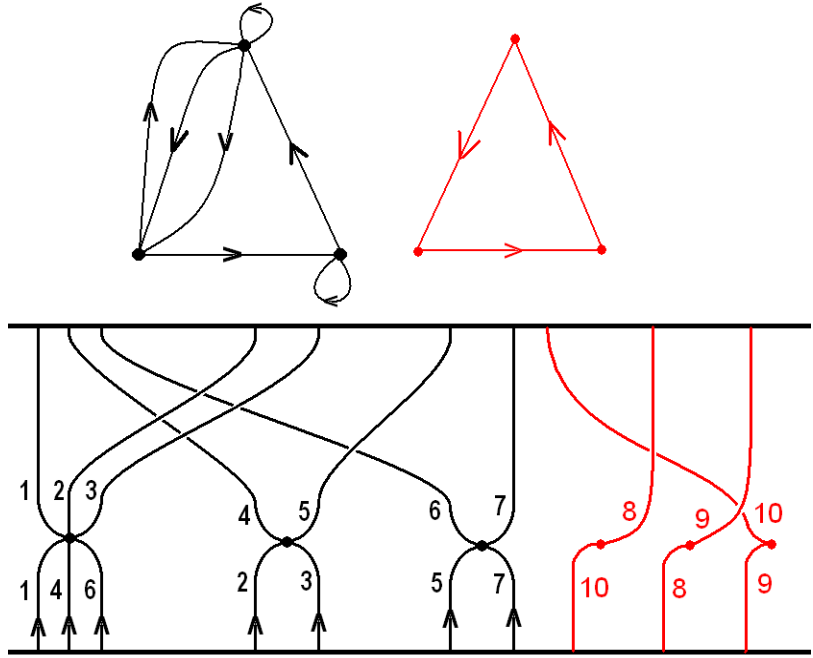}}
\caption{Any open string configuration can be mapped to a pair of labeled graphs.
         The black graph describes the $X$ matrices and the red graph the $Y$ matrices.
         The two bold horizontal lines are identified. 
         The graphs determine a permutation, so each open string configuration is mapped to a permutation. 
         For the graph shown the permutation in cycle notation is $\sigma= (2,4)(5,3,6)(8,10,9)$.
         The figure shows a configuration for a three giant system with ten open strings attached.
         Equivalently, this is an operator whose Young diagram describing the $Z$ fields has 3 long rows/columns
         and $p=7$, $m=3$.
         The vectors $\vec{p}$ and $\vec{m}$ describe the number of strings leaving each node.
         Thus, $\vec{p}=(3,2,2)$, $\vec{m}=(1,1,1)$.}
 \label{fig:gaussgraph}
\end{center}
\end{figure}

We will make use of two subgroups in what follows
\bea
   H_Y=S_{m_1}\times\cdots\times S_{m_g}\qquad H_X=S_{p_1}\times\cdots\times S_{p_g}\, .
\eea
$H_X$ acts on boxes in the partly labeled Young diagrams that are labeled with an integer $i<p+1$,
i.e. on the boxes associated to $X$s. 
$H_Y$ acts on boxes associated to $Y$s.
These two subgroups leave all partly labeled Young diagrams invariant. 
Consequently, the partly labeled Young diagrams belong to $S_p\times S_m/H_X\times H_Y$.
The Gauss graphs themselves are in one-to-one correspondence with elements of the double coset
\bea
  H_X\times H_Y \setminus S_p\times S_m /H_X\times H_Y
\label{gaussdc}
\eea
Introduce the states (these states span $V^{\otimes p+m}$)
\bea
  |v,\vec{p},\vec{m}\rangle \equiv |v_1^{\otimes p_1}\otimes v_2^{\otimes p_2}\otimes\cdots v_g^{\otimes p_g}\otimes
  v_{g+1}^{\otimes m_1}\otimes v_{g+2}^{\otimes m_2}\otimes\cdots v_{2g}^{\otimes m_g}\rangle\, .
  \label{stdform}
\eea
There is an action of the $S_p\times S_m$ group defined on this space by
\bea
  \sigma |v_{i_1}\otimes\cdots\otimes v_{i_{m+p}}\rangle = |v_{i_{\sigma (1)}}\otimes\cdots\otimes v_{i_{\sigma (m+p)}}\rangle\, .
\eea
This can trivially be enlarged to obtain an action of $S_{p+m}$, but we want to consider only permutations that mix $X$ indices
with each other and $Y$ indices with each other, but not $X$ and $Y$ indices.
Introduce the notation $|v_\sigma\rangle\equiv \sigma|v,\vec{p},\vec{m}\rangle $.
Invariance under the $H_X\times H_Y$ subgroup can be written as 
\bea
  |v_\sigma\rangle =|v_{\sigma\gamma}\rangle \quad\gamma\in H_X \times H_Y
\eea
or even
\bea
   |v_\sigma\rangle ={1\over |H_X\times H_Y|}\sum_{\gamma\in H_X\times H_Y}|v_{\sigma\gamma}\rangle\, .
\eea
Recall that the operator that projects onto representation $r$ of a group ${\cal G}$ is given by\cite{FH}
\bea
   P_r={d_r \over |{\cal G}|}\sum_{g\in{\cal G}}\chi_r(g) g\, .
\eea
By the identity representation we mean the representation for which all the elements of $H_X\times H_Y$ are represented by 1.
We want to project onto the identity representation of $H_X\times H_Y$ within the carrier space $(s,t)$ organizing the $X$s and $Y$s. 
Recall that $t\vdash p$ and $s\vdash m$. 
The characters in the identity representation are of course all equal to 1. 
The identity representation may appear more than once in $(s,t)$. 
Resolve these different copies with a multiplicity label $\vec{\mu}$.
The multiplicity label has two components, one that refers to $s$ and one that refers to $t$.
Introduce branching coefficients that resolve these projectors into a set of projectors onto each of the one dimensional spaces labeled 
by $\vec{\mu}$
\bea
  {1\over |H_X\times H_Y|}\sum_{\gamma\in H_X\times H_Y}\Gamma^{(s,t)}(\gamma)_{ik}
  =\sum_{\vec{\mu}} B^{(s,t)\to 1_{H_X\times H_Y}}_{i\vec{\mu}}B^{s\to 1_{H_X\times H_Y}}_{k\vec{\mu}}\, .
\eea
Thus, for example, $B^{(s,t)\to 1_H}_{i \vec{\nu}}B^{(s,t)\to 1_H}_{k\vec{\nu}}$ projects onto the copy $\nu_1$ of the identity representation of
$H_X$ in $s$ and onto the copy $\nu_2$ of the identity representation of $H_Y$ in $t$.
The branching coefficient $B^{(s,t)\to 1_{H_X\times H_Y}}_{i\vec{\mu}}$ can be understood as the one dimensional vector that spans the
$\vec{\nu}$ copy of $1_{H_X\times H_Y}$ inside the carrier space of $(s,t)$
\bea
  |\vec{v}\rangle_i = B_{i\vec{\nu}}^{(s,t)\to 1_{H_X\times H_Y}}\, .
\eea
Vector orthogonality says
\bea
   \langle\vec{\nu}|\vec{\mu}\rangle = \delta_{\vec{\mu}\vec{\nu}}=\sum_i B^{(s,t)\to 1_{H_X\times H_Y}}_{i\vec{\mu}}
                                                                          B^{(s,t)\to 1_{H_X\times H_Y}}_{i\vec{\nu}}
   \label{vo}
\eea
whilst vector completeness says
\bea
  \sum_{\vec{\mu}} |\vec{\mu}\rangle\langle\vec{\mu}| = 1_{H_X\times H_Y}
\eea
or, displaying all indices,
\bea
    \sum_{\vec{\mu}} B^{(s,t)\to 1_{H_X\times H_Y}}_{i\vec{\mu}} B^{(s,t)\to 1_{H_X\times H_Y}}_{j\vec{\mu}}=(1_{H_X\times H_Y})_{ij}\, .
   \label{vc}
\eea
Together (\ref{vo}) and (\ref{vc}) allow us to think of the branching coefficients 
$B^{(s,t)\to 1_{H_X\times H_Y}}_{i\vec{\mu}}$ as a matrix that implements a change of basis
\bea
   |i\rangle = \sum_{\vec{\mu}} B^{(s,t)\to 1_{H_X\times H_Y}}_{i\vec{\mu}} |\vec{\mu}\rangle\qquad
   |\vec{\mu}\rangle = \sum_{i} B^{(s,t)\to 1_{H_X\times H_Y}}_{i\vec{\mu}} |i\rangle\, .
\eea

We are now ready to argue that the Gauss operators are simply an alternative basis to the restricted Schur polynomials.
First, following \cite{deMelloKoch:2012ck}, we will show that the number of restricted Schur polynmials is equal to the number of Gauss operators.
Towards this end consider
\bea
|v_{(s,t)},i,j\rangle = \sum_{\sigma\in S_m\times S_p}\Gamma^{(s,t)}_{ij}(\sigma)\sigma |v,\vec{p},\vec{m}\rangle\, .
\eea
Above we have projected onto the representation ${(s,t)}$ of $S_m\times S_p$. 
You can think of $j$ as a label for different vectors and of $i$ as the components of the vector.
According to \cite{Koch:2011hb} this space is organized by the Schur-Weyl duality between $S_m\times S_p$ and $U(m)\times U(p)$.
Concretely, we can trade the index $j$ for a Gelfand-Tsetlin pattern.
Thus, using \cite{Koch:2011hb} we know that we can decompose this space as
\bea
V_g^{\otimes p+m}&&=\oplus_{{}^{s\vdash m\,\, t\vdash p}_{c_1(s)\le g\,\, c_1(t)\le g}}V_{(s,t)}^{U(m)\times U(p)}\otimes V_{(s,t)}^{S_m\times S_p}\cr
&&=\oplus_{{}^{s\vdash m\,\, t\vdash p}_{c_1(s)\le g\,\, c_1(t)\le g}}\oplus_{\vec{m}\,\,\vec{p}}
V_{(s,t)\to (\vec{m},\vec{p})}^{U(g)\times U(g)\to U(1)^g\times U(1)^g}\otimes V_{(s,t)}^{S_m\times S_p}\, .
\label{firstdecomp}
\eea
The first factor in the last line above is the space of Gelfand-Tsetlin patterns.
Now, lets consider a different decomposition of this space, as follows\cite{deMelloKoch:2012ck}
\bea
|v_{(s,t)},i,j\rangle 
&&=\sum_{\sigma\in S_m\times S_p}\Gamma^{(s,t)}_{ij}(\sigma)\sigma |v,\vec{p},\vec{m}\rangle\cr 
&&=\sum_{\sigma\in S_m\times S_p}\Gamma^{(s,t)}_{ij}(\sigma)|v_\sigma\rangle\cr 
&&={1\over |H_X\times H_Y|}\sum_{\sigma\in S_m\times S_p}\sum_{\gamma\in H_X\times H_Y}\Gamma^{(s,t)} (\sigma\gamma)_{ij}|v_\sigma\rangle\cr
&&={1\over |H_X\times H_Y|}\sum_{\sigma\in S_m\times S_p}\sum_{\gamma\in H_X\times H_Y}\Gamma^{(s,t)} (\sigma)_{ik}
\Gamma^{(s,t)}(\gamma)_{kj}|v_\sigma\rangle\cr
&&=\sum_{\sigma\in S_m\times S_p}\Gamma^{(s,t)} (\sigma)_{ik} \sum_{\vec{\mu}}B^{{(s,t)}\to 1_{H_X\times H_Y}}_{k\vec{\mu}}
                                                 B^{{(s,t)}\to 1_{H_X\times H_Y}}_{j\vec{\mu}}|v_\sigma\rangle\, .
\eea
As we have already discussed, the branching coefficients provide a natural change of basis from one space to the other
\bea
   |\vec{m},\vec{p},(s,t),\vec{\mu};i \rangle =\sum_j B_{j\vec{\mu}}^{(s,t)\to 1_{H_X\times H_Y}}
                                              \sum_{\sigma\in S_m\times S_p}\Gamma^{(s,t)}_{ij}(\sigma)\sigma |v_\sigma\rangle\, .
\eea
This decomposition is\cite{deMelloKoch:2012ck}
\bea
V_g^{\otimes p+m}
&&=\oplus_{{}^{s\vdash m\,\, t\vdash p}_{c_1(s)\le g\,\, c_1(t)\le g}}V_{(s,t)}^{U(m)\times U(p)}\otimes V_{(s,t)}^{S_m\times S_p}\cr
&&=\oplus_{{}^{s\vdash m\,\, t\vdash p}_{c_1(s)\le g\,\, c_1(t)\le g}}\oplus_{\vec{m}\,\,\vec{p}}
V_{(s,t)\to 1_{H_X\times H_Y}}^{S_m\times S_p\to H_X\times H_Y}\otimes V_{(s,t)}^{S_m\times S_p}\, .
\label{seconddecomp}
\eea
Comparing (\ref{firstdecomp}) to (\ref{seconddecomp}) we conclude that
\bea
   |V_{(s,t)\to (\vec{m},\vec{p})}^{U(g)\times U(g)\to U(1)^g\times U(1)^g}|=|V_{(s,t)\to 1_{H_X\times H_Y}}^{S_m\times S_p\to H_X\times H_Y}|\, .
   \label{decomp}
\eea

Using the idea that the branching coefficients provide a transformation between two bases,
we easily write the Gauss operators
\bea
   O_{R,r}(\sigma_X,\sigma_Y)&&={|H_X\times H_Y|\over\sqrt{m!p!}}\sum_{jk}\sum_{s\vdash m}\sum_{t\vdash p}\sum_{\vec{\mu}}\sum_{\vec{\nu}}
                               \sqrt{d_s d_t}\Gamma^{(s,t)}(\sigma_X\circ\sigma_Y)_{jk}\cr
                      &&\times B_{j\vec{\mu}}^{(s,t)\to 1_{H_X\times H_Y}}
                               B_{k\vec{\nu}}^{(s,t)\to 1_{H_X\times H_Y}}
                               O_{R,(t,s,r)\vec{\mu}\vec{\nu}}\, .
 \label{ggops}
\eea
Note that the factor $\sqrt{d_s d_t}$ can not be determined by group theory alone.
It is chosen so that the group theoretic coefficients
\bea
   C^{(s,t)}_{\vec{\mu}\vec{\nu}}(\sigma_X\circ\sigma_Y)
                     ={|H_X\times H_Y|\over\sqrt{m!p!}}\sum_{jk}\sum_{s\vdash m}\sum_{t\vdash p}
                      \sqrt{d_s d_t}\Gamma^{(s,t)}(\sigma_X\circ\sigma_Y)_{jk}
                      B_{j\vec{\mu}}^{(s,t)\to 1_{H_X\times H_Y}}
                      B_{k\vec{\nu}}^{(s,t)\to 1_{H_X\times H_Y}}\cr
\eea
provide an orthogonal transformation between the restricted Schur polynomials and the Gauss graph basis.
Indeed,
\bea
   \sum_{(s,t)}\sum_{\vec{\mu}}\sum_{\vec{\nu}}
    C^{(s,t)}_{\vec{\mu}\vec{\nu}}(\sigma_X\circ\sigma_Y)
    C^{(s,t)}_{\vec{\mu}\vec{\nu}}(\tau_X\circ\tau_Y) 
      =\sum_{\gamma\in H_X\times H_Y}\delta (\gamma_1\, \sigma_X\circ\sigma_Y\, \gamma_2\, \tau_X^{-1}\circ\tau_Y^{-1})\, .
\eea

There is an important point that is worth stressing here: Our Gauss operators are normalized as
\bea
   \langle O_{R,r}(\sigma_X,\sigma_Y) O_{R,r}(\tau_X,\tau_Y)^\dagger\rangle
      =\sum_{\gamma\in H_X\times H_Y}\delta (\gamma_1\, \sigma_X\circ\sigma_Y\, \gamma_2\, \tau_X^{-1}\circ\tau_Y^{-1})\, .
\eea
These operators certainly do not have unit two point function.
For example, if we set both $\sigma_X,\sigma_Y$ and $\tau_X,\tau_Y$ equal to the identity permutation, the right hand side
evaluates to $|H_X\times H_Y|$.
Our final answer is simplest when expressed in terms of normalized operators
\bea
   \hat{O}_{R,r}(\sigma_X,\sigma_Y)\equiv {1\over N_{\sigma_X,\sigma_Y}}O_{R,r}(\sigma_X,\sigma_Y)\, ,
\eea
\bea
   N_{\sigma_X,\sigma_Y}^2 =    \langle O_{R,r}(\sigma_X,\sigma_Y) O_{R,r}(\sigma_X,\sigma_Y)^\dagger\rangle\, .
\eea
We will not obtain or need the explicit form of $N_{\sigma_X,\sigma_Y}$.

\section{Dilatation Operator in the Gauss Graph Basis}

We will now write the  term in the dilatation operator that mixes $X$ and $Y$ in the Gauss graph basis, i.e. we will write this term
in the basis provided by (\ref{ggops}).
We already know that the other two terms are diagonal in this basis and we know their detailed form\cite{Koch:2011hb,deMelloKoch:2012ck}.

Towards this end, transform the intertwining operator used to construct the restricted Schur polynomial
\bea
  P_{R,(t,s,r)\vec{\mu}\vec{\nu}}=\sum_i\,\, |\vec{m},\vec{p},(s,t),\vec{\mu};i \rangle
                                           \langle \vec{m},\vec{p},(s,t),\vec{\nu};i |\otimes {\bf 1}_r
\eea
to the Gauss graph basis. Of course, it is only
\bea
  p^{\vec{m}\vec{p}}_{(t,s)\vec{\mu}\vec{\nu}}=\sum_i\,\, |\vec{m},\vec{p},(s,t),\vec{\mu};i \rangle
                                           \langle \vec{m},\vec{p},(s,t),\vec{\nu};i |
\eea
that we need to consider. The transformation is a simple computation
\bea 
&& \sum_{(s,t)} |\vec{m},\vec{p},(s,t),\vec{\mu};i\rangle \langle \vec{m},\vec{p},(s,t),\vec{\nu};i| 
B_{l\vec{\nu}}^{(s,t)\to 1_{H_X\times H_Y}} B_{m\vec{\mu}}^{(s,t)\to 1_{H_X\times H_Y}} \Gamma_{lm}^{(s,t)} (\sigma_2) \cr 
&&  ={ 1 \over |H_X\times H_Y| m! p!}   \sum_{(s,t)} \sum_{ \sigma , \tau \in S_m\times S_p } 
   d_s d_t ~~ B^{(s,t)\to 1_{H_X\times H_Y}}_{j\vec{\mu}} \Gamma^{(s,t)}_{bj} (\sigma ) |v_{\sigma} \rangle 
   \langle v_{ \tau } | \Gamma^{(s,t)}_{bk} (\tau) B^{(s,t)\to 1_{H_X\times H_Y}}_{k\vec{\nu}} \cr
&&\qquad \times B_{l\vec{\nu}}^{(s,t)\to 1_{H_X\times H_Y}} B_{m\vec{\mu}}^{(s,t)\to 1_{H_X\times H_Y}} \Gamma_{lm}^{(s,t)} (\sigma_2)\cr 
&& ={1\over |H_X\times H_Y|m!p!} \sum_{(s,t)} \sum_{\sigma ,\tau\in S_m\times S_p}  d_s d_t ~~ 
 |v_{\sigma}\rangle \langle v_{\tau}| \Gamma^{(s,t)}_{jk}(\sigma^{-1}\tau)
 B^{(s,t)\to 1_{H_X\times H_Y}}_{j\vec{\mu}} B^{(s,t)\to 1_{H_X\times H_Y}}_{k\vec{\nu}}\cr 
&&\qquad B^{(s,t)\to 1_{H_X\times H_Y}}_{l\vec{\nu}} B^{(s,t)\to 1_{H_X\times H_Y}}_{m\vec{\mu}} \Gamma_{lm}(\sigma_2)\cr 
&& = { 1 \over |H_X\times H_Y| m! p!} \sum_{(s,t)}  \sum_{ \sigma , \tau \in S_m\times S_p}  { 1 \over |H_X\times H_Y|^2 }  
    \sum_{ \gamma_1 , \gamma_2 \in H_X\times H_Y }  
 d_s d_t \Gamma_{jm}^{(s,t)} ( \gamma_1 ) \Gamma_{ kl }^{(s,t)} ( \gamma_2 ) 
\Gamma_{ lm }^{(s,t)} ( \sigma_2 ) \cr
&&\qquad\times\Gamma_{ jk }^{(s,t)} ( \sigma^{-1}  \tau ) 
| v_{ \sigma } \rangle \langle v_{ \tau } | \cr 
&& =  { 1 \over m!p!  |H_X\times H_Y| } \sum_{(s,t)} \sum_{ \sigma , \tau \in S_m\times S_p}  { 1 \over |H_X\times H_Y|^2 } 
     \sum_{ \gamma_1 , \gamma_2 \in H_X\times H_Y } 
    d_s d_t \chi_u ( \gamma_1 \sigma_2^{-1} \gamma_2^{-1} \tau^{-1} \sigma )     | v_{ \sigma } \rangle \langle v_{ \tau } |  \cr 
&& = { 1 \over |H_X\times H_Y|^3 }  \sum_{ \sigma , \tau \in S_m\times S_p} \sum_{ \gamma_1 , \gamma_2 \in H_X\times H_Y }
\delta (  \gamma_1 \sigma_2^{-1} \gamma_2^{-1} \tau^{-1} \sigma  ) 
| v_{ \sigma } \rangle \langle v_{ \tau } |\, .
\eea 
Notice that up to normalization this is a sum over all $\sigma,\tau\in S_m\times S_p$ of $|v_{\sigma}\rangle \langle v_{\tau}|$
with the condition that $\tau^{-1} \sigma$ belongs to the same coset as $\sigma_2$ does. 
With this result in hand, we easily find
\bea
\langle O^\dagger_{T,w}(\sigma_2)D_{XY}O_{R,r}(\sigma_1)\rangle\qquad\qquad=\cr
={|H_X\times H_Y||H'_X\times H'_Y|\over m!p!}\sum_{jk}\sum_{s\vdash m}\sum_{t\vdash p}\sum_{\vec{\mu}\vec{\nu}}
                                             \sum_{lm}\sum_{x\vdash m}\sum_{y\vdash p}\sum_{\vec{\alpha}\vec{\beta}}
                                             \sqrt{d_s d_t}\sqrt{d_x d_y}
                                             \Gamma^{(s,t)}(\sigma_1)_{jk}\Gamma^{(x,y)}(\sigma_2)_{lm}\cr
\times B^{(s,t)\to 1_{H_X\times H_Y}}_{j\vec{\mu}}B^{(s,t)\to 1_{H_X\times H_Y}}_{k\vec{\nu}}
       B^{(x,y)\to 1_{H'_X\times H'_Y}}_{l\vec{\alpha}}B^{(x,y)\to 1_{H'_X\times H'_Y}}_{m\vec{\beta}}
       \langle O^\dagger_{T,(y,x,w)\vec{\alpha}\vec{\beta}}D_{XY}O_{R,(t,s,r)\vec{\mu}\vec{\nu}}\rangle\cr
\cr
={|H_X\times H_Y||H'_X\times H'_Y|\over m!p!}\sum_{jk}\sum_{s\vdash m}\sum_{t\vdash p}\sum_{\vec{\mu}\vec{\nu}}
                                             \sum_{lm}\sum_{x\vdash m}\sum_{y\vdash p}\sum_{\vec{\alpha}\vec{\beta}}
                                             \sqrt{d_s d_t}\sqrt{d_x d_y}
                                             \Gamma^{(s,t)}(\sigma_1)_{jk}\Gamma^{(x,y)}(\sigma_2)_{lm}\cr
\times B^{(s,t)\to 1_{H_X\times H_Y}}_{j\vec{\mu}}B^{(s,t)\to 1_{H_X\times H_Y}}_{k\vec{\nu}}
       B^{(x,y)\to 1_{H'_X\times H'_Y}}_{l\vec{\alpha}}B^{(x,y)\to 1_{H'_X\times H'_Y}}_{m\vec{\beta}}\cr
\times\sum_{R'} c_{RR'} {d_T mp\over d_x d_y (n+m+p)d_{R'}}
    \sqrt{f_T {\rm hooks}_T{\rm hooks}_s{\rm hooks}_t\over f_R {\rm hooks}_R{\rm hooks}_x{\rm hooks}_y}\delta_{rw}\cr
\times
\Big(
  \langle\vec{p},t,\nu_1;a   |E^{(1)}_{lj}  |\vec{p}',y,\alpha_1;b\rangle
  \langle\vec{p}',y,\beta_1;b|E^{(1)}_{ki}  |\vec{p},t,\mu_1;a\rangle\cr
  \qquad\times
  \langle\vec{m},s,\nu_2;c   |E^{(p+1)}_{il}|\vec{m}',x,\alpha_2;d\rangle
  \langle\vec{m}',x,\beta_2;d|E^{(p+1)}_{jk}|\vec{m},s,\mu_2;c\rangle\cr
  -\delta_{\vec{p}\vec{p}'}\delta_{yt}\delta_{\nu_1\alpha_1}\delta_{\vec{m}\vec{m}'}\delta_{sx}\delta_{\beta_2\mu_2}
  \langle\vec{p}',y,\beta_1;b|E^{(1)}_{ji}  |\vec{p},t,\mu_1;a\rangle
  \langle\vec{m},s,\nu_2;c   |E^{(p+1)}_{ij}|\vec{m}',x,\alpha_2;d\rangle\cr
  -\delta_{\vec{p}\vec{p}'}\delta_{yt}\delta_{\mu_1\beta_1}\delta_{\vec{m}\vec{m}'}\delta_{sx}\delta_{\alpha_2\nu_2}
  \langle\vec{p},t,\nu_1;a  |E^{(1)}_{ij} |\vec{p}',y,\alpha_1;b\rangle
  \langle\vec{m}',x,\beta_2;d|E^{(p+1)}_{ji}|\vec{m},s,\mu_2;c\rangle\cr
  +\langle\vec{p},t,\nu_1;a   |E^{(1)}_{il}  |\vec{p}',y,\alpha_1;b\rangle
  \langle\vec{p}',y,\beta_1;b |E^{(1)}_{jk}|\vec{p},t,\mu_1;a\rangle\cr
  \qquad\times
  \langle\vec{m},s,\nu_2;c    |E^{(p+1)}_{lj}|\vec{m}',x,\alpha_2;d\rangle
  \langle\vec{m}',x,\beta_2;d |E^{(p+1)}_{ki}|\vec{m},s,\mu_2;c\rangle\Big)\cr
\cr
={|H_X\times H_Y||H'_X\times H'_Y|\over m!p!}\sum_{jk}\sum_{s\vdash m}\sum_{t\vdash p}\sum_{\vec{\mu}\vec{\nu}}
                                             \sum_{lm}\sum_{x\vdash m}\sum_{y\vdash p}\sum_{\vec{\alpha}\vec{\beta}}                                             
                                             \Gamma^{(s,t)}(\sigma_1)_{jk}\Gamma^{(x,y)}(\sigma_2)_{lm}\cr
\times B^{(s,t)\to 1_{H_X\times H_Y}}_{j\vec{\mu}}B^{(s,t)\to 1_{H_X\times H_Y}}_{k\vec{\nu}}
       B^{(x,y)\to 1_{H'_X\times H'_Y}}_{l\vec{\alpha}}B^{(x,y)\to 1_{H'_X\times H'_Y}}_{m\vec{\beta}}\cr
\sum_{R'} c_{RR'} {d_T mp\over (n+m+p)d_{R'}}
    \sqrt{f_T {\rm hooks}_T\over f_R {\rm hooks}_R}\delta_{rw}
\Big(
  \langle\vec{p},t,\nu_1;a   |E^{(1)}_{lj}  |\vec{p}',y,\alpha_1;b\rangle
  \langle\vec{p}',y,\beta_1;b|E^{(1)}_{ki}  |\vec{p},t,\mu_1;a\rangle\cr
  \qquad\times
  \langle\vec{m},s,\nu_2;c   |E^{(p+1)}_{il}|\vec{m}',x,\alpha_2;d\rangle
  \langle\vec{m}',x,\beta_2;d|E^{(p+1)}_{jk}|\vec{m},s,\mu_2;c\rangle\cr
  -\delta_{\vec{p}\vec{p}'}\delta_{yt}\delta_{\nu_1\alpha_1}\delta_{\vec{m}\vec{m}'}\delta_{sx}\delta_{\beta_2\mu_2}
  \langle\vec{p}',y,\beta_1;b|E^{(1)}_{ji}  |\vec{p},t,\mu_1;a\rangle
  \langle\vec{m},s,\nu_2;c   |E^{(p+1)}_{ij}|\vec{m}',x,\alpha_2;d\rangle\cr
  -\delta_{\vec{p}\vec{p}'}\delta_{yt}\delta_{\mu_1\beta_1}\delta_{\vec{m}\vec{m}'}\delta_{sx}\delta_{\alpha_2\nu_2}
  \langle\vec{p},t,\nu_1;a  |E^{(1)}_{ij} |\vec{p}',y,\alpha_1;b\rangle
  \langle\vec{m}',x,\beta_2;d|E^{(p+1)}_{ji}|\vec{m},s,\mu_2;c\rangle\cr
  +\langle\vec{p},t,\nu_1;a   |E^{(1)}_{il}  |\vec{p}',y,\alpha_1;b\rangle
  \langle\vec{p}',y,\beta_1;b |E^{(1)}_{jk}|\vec{p},t,\mu_1;a\rangle\cr
  \qquad\times
  \langle\vec{m},s,\nu_2;c    |E^{(p+1)}_{lj}|\vec{m}',x,\alpha_2;d\rangle
  \langle\vec{m}',x,\beta_2;d |E^{(p+1)}_{ki}|\vec{m},s,\mu_2;c\rangle\Big)\nn\, .
\eea
There are four terms in the above expression. We will deal with each term, one at a time.
 
\subsection{First term}

Focus on the first term for now
{\small
\bea
={|H_X\times H_Y||H'_X\times H'_Y|\over m!p!}\sum_{jk}\sum_{s\vdash m}\sum_{t\vdash p}\sum_{\vec{\mu}\vec{\nu}}
                                             \sum_{lm}\sum_{x\vdash m}\sum_{y\vdash p}\sum_{\vec{\alpha}\vec{\beta}}                                             
                                             \Gamma^{(s,t)}(\sigma_1)_{jk}\Gamma^{(x,y)}(\sigma_2)_{lm}\cr
\times B^{(s,t)\to 1_{H_X\times H_Y}}_{j\vec{\mu}}B^{(s,t)\to 1_{H_X\times H_Y}}_{k\vec{\nu}}
       B^{(x,y)\to 1_{H'_X\times H'_Y}}_{l\vec{\alpha}}B^{(x,y)\to 1_{H'_X\times H'_Y}}_{m\vec{\beta}}\cr
\sum_{R'} c_{RR'} {d_T mp\over (n+m+p)d_{R'}}
    \sqrt{f_T {\rm hooks}_T\over f_R {\rm hooks}_R}\delta_{rw}
  \langle\vec{p},t,\nu_1;a   |E^{(1)}_{lj}  |\vec{p}',y,\alpha_1;b\rangle
  \langle\vec{p}',y,\beta_1;b|E^{(1)}_{ki}  |\vec{p},t,\mu_1;a\rangle\cr
  \qquad\times
  \langle\vec{m},s,\nu_2;c   |E^{(p+1)}_{il}|\vec{m}',x,\alpha_2;d\rangle
  \langle\vec{m}',x,\beta_2;d|E^{(p+1)}_{jk}|\vec{m},s,\mu_2;c\rangle\cr
={1\over |H_X\times H_Y|^2 |H'_X\times H'_Y|^2 m!p!}\cr
\sum_{R'} c_{RR'}{\rm hooks}_{R'} mp
    \sqrt{f_T \over f_R {\rm hooks}_R {\rm hooks}_T}\delta_{rw}
\sum_{\tau\in S_m\times S_p} 
   \sum_{\phi\in S_m\times S_p}
   \sum_{\gamma_1,\gamma_2\in H_X'\times H_Y'}
   \sum_{\beta_1,\beta_2\in H_X\times H_Y}\cr
\times \langle v,\vec{p}',\vec{m}'| E^{\tau^{-1}(p+1)}_{ji}\tau^{-1} \, (1,p+1) \,\phi\beta_2\sigma_2\beta_1^{-1} |v,\vec{p},\vec{m}\rangle
\langle v,\vec{p},\vec{m}|E^{\phi^{-1}(p+1)}_{ij}\phi^{-1} \, (1,p+1)\, \tau\gamma_2\sigma_1\gamma_1^{-1}|v,\vec{p}',\vec{m}'\rangle\nn
\eea}
Now, lets study the case that $i=j$.
To find a simple condition on $\vec{p}',\vec{m}'$ and $\vec{p},\vec{m}$ that tells us when this matrix element is non-zero, focus on
\bea
   \langle v_{\tau}|E^{(p+1)}_{ji} \, (1,p+1) \,|v_{\psi} \rangle \langle v_{\phi}|\, (1,p+1)\,E^{(1)}_{ij}|v_{\sigma} \rangle\, .
\eea

If $i=j$, the matrix element $\langle v,\vec{p}',\vec{m}'|\tau^{-1}E^{(p+1)}_{ji} \, (1,p+1) \,\psi|v,\vec{p},\vec{m} \rangle$
forces $\vec{p}+\vec{m}=\vec{p}'+\vec{m}'$. 
Indeed, $E^{(p+1)}_{ii}$ is one if the vector in the first slot of $\psi|v,\vec{p},\vec{m}\rangle$
is $v_1$ and it is zero otherwise, so it clearly does not change the identity of any vectors. 
The remaining elements between the two states (i.e. $\tau^{-1}$ and $(1,p+1) \,\psi$) can swap vectors around but not change the identity 
of any vector. 
Thus, the identity of the collection of vectors used to construct $|v,\vec{p},\vec{m}\rangle$ must match the identity of the collection of 
vectors used to construct $|v,\vec{p}',\vec{m}'\rangle$. 
This then proves that $\vec{p}+\vec{m}=\vec{p}'+\vec{m}'$.
We can argue for this conclusion in a second way: recall that we obtain $R'$ from $R$ by dropping box in row $i$ and we obtain $T'$ from $T$
by dropping a box in row $j$.
Thus, if $i=j$, since $R'=T'$ we are saying that $R=T$. 
We already know that $r=w$.
$\vec{p}+\vec{m}$ tells us the collection of boxes that needs to be dropped from $R$ to get $r$ and
$\vec{p}'+\vec{m}'$ tells us the collection of boxes that needs to be dropped from $T$ to get $w$.
Since $R=T$ and $r=w$, this then again proves that $\vec{p}+\vec{m}=\vec{p}'+\vec{m}'$. 
We can say a bit more. Consider
\bea
   \langle v,\vec{p},\vec{m}|\phi^{-1}\, (1,p+1)\,E^{(1)}_{ii}\sigma |v,\vec{p}',\vec{m}'\rangle\, .
\eea
This tells you that if you take the state $|v,\vec{p}',\vec{m}'\rangle$ and shuffle some of the $X$ slots amongst
each other and some of the $Y$ slots amongst each other ($\sigma$ does this shuffling) keeping only states with
vector $v_i$ in their first slot, and then swapping the vectors in slots $1$ and $p+1$, we can get the vector
$|v,\vec{p},\vec{m}\rangle$ by shuffling (according to $\phi^{-1}$) what we have. {\bf Thus, to get 
$|v,\vec{p},\vec{m}\rangle$ from $|v,\vec{p}',\vec{m}'\rangle$ we removed $v_i$ from an $X$ slot of $|v,\vec{p}',\vec{m}'\rangle$
and inserted it into a $Y$ slot of $|v,\vec{p},\vec{m}\rangle$.}

Now consider
\bea
   \langle v,\vec{p}',\vec{m}'|\tau^{-1}E^{(p+1)}_{ii} \, (1,p+1) \psi|v,\vec{p},\vec{m} \rangle\, .
\eea
This tells you that if you take the state $|v,\vec{p},\vec{m}\rangle$ and shuffle some of the $X$ slots amongst
each other and some of the $Y$ slots amongst each other ($\psi$ does this shuffling) keeping only states with
vector $v_i$ in their first slot, and then, swapping the vectors in slots $1$ and $p+1$, we can get the vector
$|v,\vec{p}',\vec{m}'\rangle$ by shuffling (according to $\tau^{-1}$) what we have. {\bf Thus, to get 
$|v,\vec{p}',\vec{m}'\rangle$ from $|v,\vec{p},\vec{m}\rangle$ we removed $v_i$ from an $X$ slot of $|v,\vec{p},\vec{m}\rangle$
and inserted it into a $Y$ slot of $|v,\vec{p},\vec{m}\rangle$.}

Thus, the two vectors we are swapping have the {\it same} identity. 
This implies that we must have $\vec{p}=\vec{p}'$ and $\vec{m}=\vec{m}'$.
Since we must have $\vec{p}=\vec{p}'$ and $\vec{m}=\vec{m}'$ we find
{\small
\bea
={1\over |H_X\times H_Y|^4 m!p!}
\sum_{R'} c_{RR'}{\rm hooks}_{R'} mp
    \sqrt{f_T \over f_R {\rm hooks}_R {\rm hooks}_T}\delta_{rw}
\sum_{\tau\in S_m\times S_p} 
   \sum_{\phi\in S_m\times S_p}
   \sum_{\gamma_1,\gamma_2,\beta_1,\beta_2\in H_X\times H_Y}\cr
\times \langle v,\vec{p},\vec{m}| E^{\tau^{-1}(p+1)}_{ji}\tau^{-1} \, (1,p+1) \,\phi\beta_2\sigma_2\beta_1^{-1} |v,\vec{p},\vec{m}\rangle
\langle v,\vec{p},\vec{m}|E^{\phi^{-1}(p+1)}_{ij}\phi^{-1} \, (1,p+1)\, \tau\gamma_2\sigma_1\gamma_1^{-1}|v,\vec{p},\vec{m}\rangle\cr
={1\over |H_X\times H_Y|^4 m!p!}
\sum_{R'} c_{RR'}{\rm hooks}_{R'} mp
    \sqrt{f_T \over f_R {\rm hooks}_R {\rm hooks}_T}\delta_{rw}
\sum_{\tau\in S_m\times S_p} 
   \sum_{\phi\in S_m\times S_p}
   \sum_{\gamma_1,\gamma_2,\beta_1,\beta_2,\rho_1,\rho_2\in H_X\times H_Y}\cr
\times \delta (\tau^{-1} \,\phi\beta_2\sigma_2\beta_1^{-1}\rho_1)\delta (\phi^{-1}\, \tau\gamma_2\sigma_1\gamma_1^{-1}\rho_2)\cr
\times \sum_{k,q\in S_{i,m}}\sum_{l,r\in S_{i,p}}
\delta(\tau^{-1}(p+1),k)\delta(\tau^{-1}(1),l)\delta(\phi^{-1}(p+1),q)\delta(\phi^{-1}(1),r)\nn\, .
\eea}
Now, set $\tau =\alpha\tilde{\tau}$ and $\beta=\alpha\tilde{\beta}$ with $\alpha\in Z_{m}\times Z_{p}$, with $Z_m\times Z_p$ a product 
of cyclic groups.
The above expression becomes
{\small
\bea
={1\over |H_X\times H_Y|^4 m!p!}\sum_{R'} c_{RR'}{\rm hooks}_{R'} mp\sqrt{f_T \over f_R {\rm hooks}_R {\rm hooks}_T}\delta_{rw}
\sum_{\tau\in S_m\times S_p} \sum_{\phi\in S_m\times S_p}\cr
  \times\sum_{\gamma_1,\gamma_2,\beta_1,\beta_2,\rho_1,\rho_2\in H_X\times H_Y} \delta (\tau^{-1} 
  \,\phi\beta_2\sigma_2\beta_1^{-1}\rho_1)\delta (\phi^{-1}\, \tau\gamma_2\sigma_1\gamma_1^{-1}\rho_2)\cr
\times \sum_{k,q\in S_{i,m}}\sum_{l,r\in S_{i,p}} \delta(\tau^{-1}(\a(p+1)),k)\delta(\tau^{-1}(\a(1)),l)
\delta(\phi^{-1}(\a(p+1)),q)\delta(\phi^{-1}(\a(1)),r)\cr
=\sum_{\a\in Z_{m}\times Z_{p}}{1\over |H_X\times H_Y|^4 m!p!}
   \sum_{R'} c_{RR'}{\rm hooks}_{R'} \sqrt{f_T \over f_R {\rm hooks}_R {\rm hooks}_T}\delta_{rw}
   \sum_{\tau\in S_m\times S_p} \sum_{\phi\in S_m\times S_p}\cr
   \times\sum_{\gamma_1,\gamma_2,\beta_1,\beta_2,\rho_1,\rho_2\in H_X\times H_Y}
   \delta (\tau^{-1} \,\phi\beta_2\sigma_2\beta_1^{-1}\rho_1)\delta (\phi^{-1}\, \tau\gamma_2\sigma_1\gamma_1^{-1}\rho_2)\cr
   \times \sum_{k,q\in S_{i,m}}\sum_{l,r\in S_{i,p}}
   \delta(\tau^{-1}(\a(p+1)),k)\delta(\tau^{-1}(\a(1)),l)\delta(\phi^{-1}(\a(p+1)),q)\delta(\phi^{-1}(\a(1)),r)\cr
={1\over |H_X\times H_Y|^4 m!p!}\sum_{R'} c_{RR'}{\rm hooks}_{R'} \sqrt{f_T \over f_R {\rm hooks}_R {\rm hooks}_T}\delta_{rw}
   \sum_{\tau\in S_m\times S_p} \sum_{\phi\in S_m\times S_p}\cr
   \times\sum_{\gamma_1,\gamma_2,\beta_1,\beta_2,\rho_1,\rho_2\in H_X\times H_Y}
   \delta (\tau^{-1} \,\phi\beta_2\sigma_2\beta_1^{-1}\rho_1)\delta (\phi^{-1}\, \tau\gamma_2\sigma_1\gamma_1^{-1}\rho_2)\cr
   \times \sum_{k,q\in S_{i,m}}\sum_{l,r\in S_{i,p}}
   \delta(\tau^{-1}(\phi(q)),k)\delta(\tau^{-1}(\phi(r)),l)\cr
={1\over |H_X\times H_Y|^4} \sum_{R'} c_{RR'}{\rm hooks}_{R'} \sqrt{f_T \over f_R {\rm hooks}_R {\rm hooks}_T}\delta_{rw} 
   \sum_{\phi\in S_m\times S_p}\cr
   \times\sum_{\gamma_1,\gamma_2,\beta_1,\beta_2,\rho_1,\rho_2\in H_X\times H_Y}
   \delta (\phi\beta_2\sigma_2\beta_1^{-1}\rho_1)\delta (\phi^{-1}\, \gamma_2\sigma_1\gamma_1^{-1}\rho_2)
   \sum_{k,q\in S_{i,m}}\sum_{l,r\in S_{i,p}}\delta(\phi(q),k)\delta(\phi(r),l)\cr
={1\over |H_X\times H_Y|^2}
\sum_{R'} c_{RR'}{\rm hooks}_{R'} \sqrt{f_T \over f_R {\rm hooks}_R {\rm hooks}_T}\delta_{rw} 
   \sum_{\phi\in S_m\times S_p}
   \sum_{\gamma_1,\gamma_2,\beta_1,\beta_2\in H_X\times H_Y}\cr
\times \delta (\phi\beta_2\sigma_2\beta_1^{-1})\delta (\phi^{-1}\, \gamma_2\sigma_1\gamma_1^{-1})
\sum_{k,q\in S_{i,m}}\sum_{l,r\in S_{i,p}}
\delta(\phi(q),k)\delta(\phi(r),l)\cr
={1\over |H_X\times H_Y|^2}
\sum_{R'} c_{RR'}{\rm hooks}_{R'} \sqrt{f_T \over f_R {\rm hooks}_R {\rm hooks}_T}\delta_{rw} 
     \sum_{\gamma_1,\gamma_2,\beta_1,\beta_2\in H_X\times H_Y}\cr
\times \delta (\gamma_2\sigma_1\gamma_1^{-1}\beta_2\sigma_2\beta_1^{-1})
n_{ii}^X(\sigma_1)n_{ii}^Y(\sigma_1)\cr
=\sum_{R'} c_{RR'}{\rm hooks}_{R'} \sqrt{f_T \over f_R {\rm hooks}_R {\rm hooks}_T}\delta_{rw} 
     \sum_{\beta_1,\beta_2\in H_X\times H_Y}
    \delta (\sigma_1\beta_2\sigma_2\beta_1^{-1})n_{ii}^X(\sigma_1)n_{ii}^Y(\sigma_1)\, .\cr
\eea}

Now, return to the case that $i\ne j$.
The matrix element $\langle v,\vec{p}',\vec{m}'|\tau^{-1}E^{(p+1)}_{ji}\,(1,p+1)\,\psi|v,\vec{p},\vec{m}\rangle$ 
forces $\vec{p}+\vec{m}\ne\vec{p}'+\vec{m}'$.
Indeed, $(1,p+1) \,\psi$ shuffles vectors, $E^{(p+1)}_{ji}$ removes $v_i$ and inserts $v_j$ and $\tau^{-1}$ does some more shuffling.
Thus, using an obvious notation, we have
\bea
  \vec{p}+\vec{m}-\hat{i}=\vec{p}'+\vec{m}'-\hat{j}\, .
\eea
We can also see this by noting that since $i\ne j$ we know that $R\ne T$. 
We still have $r=w$ so that the collection of boxes that needs to be dropped from $R$ to get $r$ (described by $\vec{p}+\vec{m}$) and
the collection of boxes that needs to be dropped from $T$ to get $w$ (described by $\vec{p}'+\vec{m}'$) can't possibly be equal.

Again, we can say more. Consider
\bea
   \langle v,\vec{p},\vec{m}|\phi^{-1}\, (1,p+1)\,E^{(1)}_{ij}\sigma |v,\vec{p}',\vec{m}'\rangle\, .
\eea
This tells you that if you take the state $|v,\vec{p}',\vec{m}'\rangle$ and shuffle some of the $X$ slots amongst
each other and some of the $Y$ slots amongst each other ($\sigma$ does this shuffling) keeping only states with
vector $v_j$ in their first slot, replacing this vector $v_j$ with another vector $v_i$ and then swapping the 
vectors in slots $1$ and $p+1$, we can get the vector $|v,\vec{p},\vec{m}\rangle$ by shuffling (according to $\phi^{-1}$) 
what we have. We can summarize this as
\bea
  &&\vec{p}'-\hat{j}=\vec{p}-\hat{a}\cr
  &&\vec{m}'-\hat{a}=\vec{m}-\hat{i}\, .
  \label{fstway}
\eea

Now consider
\bea
   \langle v,\vec{p}',\vec{m}'|\tau^{-1}E^{(p+1)}_{ji} \, (1,p+1) \psi|v,\vec{p},\vec{m} \rangle\, .
\eea
This tells you that if you take the state $|v,\vec{p},\vec{m}\rangle$ and shuffle some of the $X$ slots amongst
each other and some of the $Y$ slots amongst each other ($\psi$ does this shuffling) keeping only states with
vector $v_i$ in their first slot, replacing this vector $v_i$ with $v_j$ and then, swapping the vectors in slots 
$1$ and $p+1$, we can get the vector $|v,\vec{p}',\vec{m}'\rangle$ by shuffling (according to $\tau^{-1}$) what we have. 
We can summarize this as
\bea
  &&\vec{p}-\hat{i}=\vec{p}'-\hat{b}\cr
  &&\vec{m}-\hat{b}=\vec{m}'-\hat{j}\, .
  \label{scndway}
\eea

The equations (\ref{fstway}) and (\ref{scndway}) only have two solutions. 
If we choose $\hat{a}=\hat{i}$, we must have $\hat{b}=\hat{j}$ and then
\bea
   &&\vec{m}=\vec{m}'\cr
   &&\vec{p}-\hat{i}=\vec{p}'-\hat{j}\, .
   \label{firstsolution}
\eea
If we choose $\hat{a}=\hat{j}$, we must have $\hat{b}=\hat{i}$ and then
\bea
   &&\vec{p}=\vec{p}'\cr
   &&\vec{m}-\hat{i}=\vec{m}'-\hat{j}\, .
   \label{secondsolution}
\eea
Thus, only $\vec{m}$ or $\vec{p}$ can change - but not both. 
In fact, only one of the Gauss graphs (there is one graph for the $X$s and one for the $Y$s) change - but not both.

It is now rather simple to write the relation between $|v,\vec{p}',\vec{m}'\rangle$ and $|v,\vec{p},\vec{m}\rangle$.
Consider for example, (\ref{firstsolution}).
Let $S_{j,p}$ denote the collection of slots that (i) are $X$ slots and (ii) are occupied by $v_j$. 
There are similar definitions for $S_{j,m}$, $S'_{j,p}$ and $S'_{j,m}$.
To go from $\vec{p}$ to $\vec{p}'$, we want to remove a $v_i$ and replace it with a $v_j$ and then reorder the slots
into the order prescribed by (\ref{stdform}). We can do this as
\bea
   |v,\vec{p}',\vec{m}'\rangle=\zeta E^{(q)}_{ji}|v,\vec{p},\vec{m}\rangle\qquad q\in S_{i,p}\quad \zeta\in S_m\times S_p\, .
\eea
Consequently we can again write a definite relation between $|v,\vec{p}',\vec{m}'\rangle$ and $|v,\vec{p},\vec{m}\rangle$.
This allows us to simplify the matrix element expressions to the structure of elements we have already evaluated.
Now, consider
\bea
A={1\over |H_X\times H_Y|^2 |H'_X\times H'_Y|^2 m!p!}\sum_{R'} c_{RR'}{\rm hooks}_{R'} mp\cr
  \sqrt{f_T \over f_R {\rm hooks}_R {\rm hooks}_T}\delta_{rw}\sum_{\tau\in S_m\times S_p} 
  \sum_{\phi\in S_m\times S_p}\sum_{\gamma_1,\gamma_2\in H_X'\times H_Y'}\sum_{\beta_1,\beta_2\in H_X\times H_Y}\cr
  \times \langle v,\vec{p}',\vec{m}'| E^{\tau^{-1}(p+1)}_{ji}\tau^{-1} \, (1,p+1) \,\phi\beta_2\sigma_2\beta_1^{-1} |v,\vec{p},\vec{m}\rangle\cr
  \times
   \langle v,\vec{p},\vec{m}|E^{\phi^{-1}(p+1)}_{ij}\phi^{-1} \, (1,p+1)\, \tau\gamma_2\sigma_1\gamma_1^{-1}|v,\vec{p}',\vec{m}'\rangle\cr
  ={1\over |H_X\times H_Y|^2 |H'_X\times H'_Y|^2 m!p!}\sum_{R'} \sqrt{c_{RR'}c_{TR'}}\cr
  \times {{\rm hooks}_{R'}\over \sqrt{{\rm hooks}_R {\rm hooks}_T}} mp \delta_{rw}
  \sum_{\tau,\phi\in S_m\times S_p}\sum_{\gamma_1,\gamma_2\in H_X'\times H_Y'}\sum_{\beta_1,\beta_2\in H_X\times H_Y}\cr
\times \langle v,\vec{p}',\vec{m}'| E^{\tau^{-1}(p+1)}_{ji}\tau^{-1} \, (1,p+1) \,\phi\beta_2\sigma_2\beta_1^{-1} |v,\vec{p},\vec{m}\rangle\cr
\times 
\langle v,\vec{p},\vec{m}|E^{\phi^{-1}(p+1)}_{ij}\phi^{-1} \, (1,p+1)\, \tau\gamma_2\sigma_1\gamma_1^{-1}|v,\vec{p}',\vec{m}'\rangle\, .
\eea
To start, study
\bea
   \langle v,\vec{p}',\vec{m}'| E^{\tau^{-1}(p+1)}_{ji}\tau^{-1} \, (1,p+1) \,\phi\beta_2\sigma_2\beta_1^{-1} |v,\vec{p},\vec{m}\rangle\cr
=   \langle v,\vec{p}',\vec{m}'|\tau^{-1} E^{(p+1)}_{ji} \, (1,p+1) \,\phi\beta_2\sigma_2\beta_1^{-1} |v,\vec{p},\vec{m}\rangle
\eea
and consider a matrix element for which $\vec{m}=\vec{m}'$ and $\vec{p}=\vec{p}'-\hat{j}+\hat{i}$.
In this case, we know that we can write
\bea
| v,\vec{p}',\vec{m}'\rangle =\zeta E^{(q)}_{ji}|v,\vec{p},\vec{m}\rangle\qquad \zeta\in S_p \qquad q\in S_{i,p}
\eea
We can choose any basis for the vectors $|v,\vec{p},\vec{m}\rangle$, $|v,\vec{p}',\vec{m}'\rangle$ that we like - the result will be independent of the choice we make.
In (\ref{stdform}) choose the $i$ and $j$ vectors to sit in adjacent slots, and always choose $q$ to lie on the border between the two.
In this case we can always choose $\zeta_q$ to be the identity.
With this choice understood, we have
\bea
| v,\vec{p}',\vec{m}'\rangle = E^{(q)}_{ji}|v,\vec{p},\vec{m}\rangle\qquad  q\in S_{i,p}\, .
\label{relatevectors}
\eea
In a similar way
\bea
   \langle v,\vec{p},\vec{m}|E^{\phi^{-1}(p+1)}_{ij}\phi^{-1} \, (1,p+1)\, \tau\gamma_2\sigma_1\gamma_1^{-1}|v,\vec{p}',\vec{m}'\rangle\cr
=  \langle v,\vec{p},\vec{m}|\phi^{-1}E^{(p+1)}_{ij} \, (1,p+1)\, \tau\gamma_2\sigma_1\gamma_1^{-1}|v,\vec{p}',\vec{m}'\rangle
\eea
and, from (\ref{relatevectors}) we have
\bea
| v,\vec{p},\vec{m}\rangle = E^{(q)}_{ij}|v,\vec{p}',\vec{m}'\rangle\, .
\eea
Consequently
\bea
A={1\over |H_X\times H_Y|^2 |H'_X\times H'_Y|^2 m!p!}
  \sum_{R'} \sqrt{c_{RR'}c_{TR'}}{{\rm hooks}_{R'}\over \sqrt{{\rm hooks}_R {\rm hooks}_T}} mp \delta_{rw}
  \sum_{\gamma_1,\gamma_2\in H_X'\times H_Y'}\cr
  \times \sum_{\beta_1,\beta_2\in H_X\times H_Y}\sum_{\tau,\phi\in S_m\times S_p} 
  \langle v,\vec{p}',\vec{m}'| \tau^{-1} E^{(p+1)}_{ji} \, (1,p+1) \,\phi\beta_2\sigma_2\beta_1^{-1} |v,\vec{p},\vec{m}\rangle\cr
  \times \langle v,\vec{p},\vec{m}|\phi^{-1}E^{(p+1)}_{ij} \, (1,p+1)\, \tau\gamma_2\sigma_1\gamma_1^{-1}|v,\vec{p}',\vec{m}'\rangle\cr
 ={1\over |H_X\times H_Y|^2 |H'_X\times H'_Y|^2 m!p!}
  \sum_{R'} \sqrt{c_{RR'}c_{TR'}}{{\rm hooks}_{R'}\over \sqrt{{\rm hooks}_R {\rm hooks}_T}} mp \delta_{rw} 
  \sum_{\gamma_1,\gamma_2\in H_X'\times H_Y'} \cr
  \times  \sum_{\beta_1,\beta_2\in H_X\times H_Y} \sum_{\tau,\phi\in S_m\times S_p}  
  \langle v,\vec{p},\vec{m}|E^{(q)}_{ij}\tau^{-1} E^{(p+1)}_{ji} \, (1,p+1) \,\phi\beta_2\sigma_2\beta_1^{-1} |v,\vec{p},\vec{m}\rangle\cr
  \times \langle v,\vec{p}',\vec{m}'|E^{(q)}_{ji}\phi^{-1}E^{(p+1)}_{ij} \, (1,p+1)\, \tau\gamma_2\sigma_1\gamma_1^{-1}|v,\vec{p}',\vec{m}'\rangle\cr
={1\over |H_X\times H_Y|^2 |H'_X\times H'_Y|^2 m!p!}\sum_{R'}\sqrt{c_{RR'}c_{TR'}}{{\rm hooks}_{R'}\over\sqrt{{\rm hooks}_R{\rm hooks}_T}}mp\delta_{rw}
  \sum_{\gamma_1,\gamma_2\in H_X'\times H_Y'}\cr
 \times\sum_{\beta_1,\beta_2\in H_X\times H_Y}\sum_{\tau,\phi\in S_m\times S_p}  
 \langle v,\vec{p},\vec{m}|\tau^{-1} E^{\tau(q)}_{ij}E^{(p+1)}_{ji} \, (1,p+1) \,\phi\beta_2\sigma_2\beta_1^{-1} |v,\vec{p},\vec{m}\rangle\cr
 \times \langle v,\vec{p}',\vec{m}'|\phi^{-1}E^{\phi(q)}_{ji}E^{(p+1)}_{ij} \, (1,p+1)\, \tau\gamma_2\sigma_1\gamma_1^{-1}|v,\vec{p}',\vec{m}'\rangle\cr
={1\over |H_X\times H_Y|^2 |H'_X\times H'_Y|^2 m!p!}
    \sum_{R'} \sqrt{c_{RR'}c_{TR'}}{{\rm hooks}_{R'}\over\sqrt{{\rm hooks}_R {\rm hooks}_T}}mp\delta_{rw}
    \sum_{\gamma_1,\gamma_2\in H_X'\times H_Y'}\cr
    \sum_{\beta_1,\beta_2\in H_X\times H_Y}\sum_{\tau,\phi\in S_m\times S_p}  
    \langle v_{\tau}|E^{\tau(q)}_{ij}E^{(p+1)}_{ji}\, (1,p+1)\,\phi\beta_2\sigma_2\beta_1^{-1}\tau^{-1}|v_\tau\rangle)\cr
\times \langle v'_{\phi}|E^{\phi(q)}_{ji}E^{(p+1)}_{ij}\, (1,p+1)\,\tau\gamma_2\sigma_1\gamma_1^{-1}\phi^{-1}|v'_{\phi}\rangle\cr
\label{expressionsofar}
\eea
We need to understand $\langle v_{\tau}|E^{\tau(q)}_{ij}E^{(p+1)}_{ji} \, (1,p+1)$ and 
$\langle v'_{\phi}|E^{\phi(q)}_{ji}E^{(p+1)}_{ij} \, (1,p+1)$ better.
Consider $\langle v_{\tau}|E^{\tau(q)}_{ij}E^{(p+1)}_{ji} \, (1,p+1)$ first. Turn this into a ket state
\bea
(1,p+1) \, E^{(p+1)}_{ij}E^{\tau(q)}_{ji}\tau|v\rangle
=(1,p+1) \, \tau \, E^{\tau^{-1}(p+1)}_{ij}E^{(q)}_{ji}|v\rangle\cr
=\tau \, (\tau^{-1}(1),\tau^{-1}(p+1)) \, E^{\tau^{-1}(p+1)}_{ij}E^{(q)}_{ji}|v\rangle
\eea
Now, there are a few things we should note. 
First, recall that $i\ne j$.
In the matrix element
\bea
   \langle v_{\tau}|E^{\tau(q)}_{ij}E^{(p+1)}_{ji} \, (1,p+1) \,\phi\beta_2\sigma_2\beta_1^{-1} \tau^{-1}|v_\tau\rangle
\eea
we know that $\phi\beta_2\sigma_2\beta_1^{-1} \tau^{-1}$ is an element in $S_m\times S_p$ and thus it is not able to swap
vectors between the $Y$ and $X$ slots. The product $E^{\tau(q)}_{ij}E^{(p+1)}_{ji}$ makes an $X$ slot change as
(imagine acting to the right) $j\to i$ and a $Y$ slot change as $i\to j$. This amounts to exchanging
an $i$ vector from $X$ with a $j$ vector from $Y$. The only way that the above matrix element can be non-zero, is if $(1,p+1)$
is able to swap these two back again. Thus, we can write
\bea
\tau \, (\tau^{-1}(1),\tau^{-1}(p+1)) \, E^{\tau^{-1}(p+1)}_{ij}E^{(q)}_{ji}|v\rangle
=\tau \, E^{\tau^{-1}(1)}_{ij}E^{\tau^{-1}(p+1)}_{ji} \, E^{\tau^{-1}(p+1)}_{ij}E^{(q)}_{ji}|v\rangle\cr
=\sum_{l\in S_{j,m}}\delta(\tau^{-1}(p+1),l) \tau \, E^{\tau^{-1}(1)}_{ij}E^{(q)}_{ji}|v\rangle\cr
=\sum_{l\in S_{j,m}}\delta(\tau^{-1}(p+1),l) \left(
\delta(\tau^{-1}(1),q)+\sum_{r\in S_{j,p}}\delta (\tau^{-1}(1),r)\, (q,r)\,
\right)\, \tau \, |v\rangle\, .
\eea
Now, each of the terms in round brackets for which index $r$ belongs to a string that loops back to node $j$
above makes the same contribution so that we have
\bea
(1,p+1) \, E^{(p+1)}_{ij}E^{\tau(q)}_{ji}|v_{\tau}\rangle
=n_{jj}^X(\sigma_2)\sum_{l\in S_{j,m}}\delta(\tau^{-1}(p+1),l)
\delta(\tau^{-1}(1),q) |v_\tau\rangle\, .
\eea
The above equation is not exactly true (certain terms on the RHS have been dropped) but it gives the correct result when plugged 
into (\ref{expressionsofar}).
A very similar argument implies that we can replace
\bea
 (1,p+1) \, E^{(p+1)}_{ji}E^{(\phi\zeta_q)(q)}_{ij}|v'_{\phi}\rangle =n_{ii}^X(\sigma_1)
 \sum_{w\in S'_{i,m}}\delta(\phi^{-1}(p+1),w)\delta(\phi^{-1}(1),\zeta_q(q))|v'_\phi\rangle
\eea
We can now use these results to compute
\bea
  A={n_{ii}^X(\sigma_1)n_{jj}^X(\sigma_2)\over |H_X\times H_Y|^2 |H'_X\times H'_Y|^2 m!p!}
    \sum_{R'} \sqrt{c_{RR'}c_{TR'}}{{\rm hooks}_{R'}\over \sqrt{{\rm hooks}_R {\rm hooks}_T}} mp\delta_{rw}\cr
    \times\sum_{\gamma_1,\gamma_2\in H_X'\times H_Y'}\sum_{\beta_1,\beta_2\in H_X\times H_Y}\sum_{\tau\in S_m\times S_p} 
    \sum_{\phi\in S_m\times S_p}
    \sum_{l\in S_{j,m}} \sum_{w\in S'_{i,m}}\delta(\tau^{-1}(p+1),l)\delta(\phi^{-1}(p+1),w)\cr
    \delta(\tau^{-1}(1),q)\delta(\phi^{-1}(1),\zeta_q(q))
    \langle v_{\tau}|\phi\beta_2\sigma_2\beta_1^{-1}\tau^{-1}|v_\tau\rangle\,
    \langle v'_{\phi}|\tau\gamma_2\sigma_1\gamma_1^{-1} \phi^{-1}|v'_{\phi}\rangle\cr
   ={n_{ii}^X(\sigma_1)n_{jj}^X(\sigma_2)\over |H_X\times H_Y|^2 |H'_X\times H'_Y|^2 m!p!}
    \sum_{R'} \sqrt{c_{RR'}c_{TR'}}{{\rm hooks}_{R'}\over \sqrt{{\rm hooks}_R {\rm hooks}_T}} mp\delta_{rw}\cr
    \times\sum_{\gamma_1,\gamma_2,\gamma\in H_X'\times H_Y'}\sum_{\beta_1,\beta_2,\beta\in H_X\times H_Y}\sum_{\tau\in S_m\times S_p} 
    \sum_{\phi\in S_m\times S_p}
    \sum_{l\in S_{j,m}} \sum_{w\in S'_{i,m}}\delta(\tau^{-1}(p+1),l)\delta(\phi^{-1}(p+1),w)\cr
    \delta(\tau^{-1}(1),q)\delta(\phi^{-1}(1),q)
    \delta(\tau^{-1}\phi\beta_2\sigma_2\beta_1^{-1}\beta)\,
    \delta(\phi^{-1}\tau\gamma_2\sigma_1\gamma_1^{-1}\gamma)\cr
   ={n_{ii}^X(\sigma_1)n_{jj}^X(\sigma_2)\over |H_X\times H_Y| |H'_X\times H'_Y| m!p!}
    \sum_{R'} \sqrt{c_{RR'}c_{TR'}}{{\rm hooks}_{R'}\over \sqrt{{\rm hooks}_R {\rm hooks}_T}} mp\delta_{rw}\cr
    \times\sum_{\gamma_1,\gamma_2\in H_X'\times H_Y'}\sum_{\beta_1,\beta_2\in H_X\times H_Y}\sum_{\tau\in S_m\times S_p} 
    \sum_{\phi\in S_m\times S_p}
    \sum_{l\in S_{j,m}} \sum_{w\in S'_{i,m}}\delta(\tau^{-1}(p+1),l)\delta(\phi^{-1}(p+1),w)\cr
    \delta(\tau^{-1}(1),q)\delta(\phi^{-1}(1),q)
    \delta(\tau^{-1}\phi\beta_2\sigma_2\beta_1^{-1})\,
    \delta(\phi^{-1}\tau\gamma_2\sigma_1\gamma_1^{-1})\cr
\eea
Now, we do the same trick as before, setting $\phi = \alpha\tilde{\phi}$ and $\tau = \alpha\tilde{\tau}$. 
After performing manipulations  just like we did before we find
\bea
A ={n_{ii}^X(\sigma_1)n_{jj}^X(\sigma_2)\over |H_X\times H_Y| |H'_X\times H'_Y| m!p!}
    \sum_{R'} \sqrt{c_{RR'}c_{TR'}}{{\rm hooks}_{R'}\over \sqrt{{\rm hooks}_R {\rm hooks}_T}}\delta_{rw}\cr
    \times\sum_{\gamma_1,\gamma_2\in H_X'\times H_Y'}\sum_{\beta_1,\beta_2\in H_X\times H_Y}\sum_{\tau\in S_m\times S_p} 
    \sum_{\phi\in S_m\times S_p}\cr
    \times \sum_{l\in S_{j,m}} \sum_{w\in S'_{i,m}}\delta(\tau^{-1}\phi(w),l)\delta(\tau^{-1}\phi(q),q)
    \delta(\tau^{-1}\phi\beta_2\sigma_2\beta_1^{-1})\,
    \delta(\phi^{-1}\tau\gamma_2\sigma_1\gamma_1^{-1})\cr
  ={n_{ii}^X(\sigma_1)n_{jj}^X(\sigma_2)\over |H_X\times H_Y| |H'_X\times H'_Y|}
    \sum_{R'} \sqrt{c_{RR'}c_{TR'}}{{\rm hooks}_{R'}\over \sqrt{{\rm hooks}_R {\rm hooks}_T}}\delta_{rw}
    \sum_{\gamma_1,\gamma_2\in H_X'\times H_Y'}\sum_{\beta_1,\beta_2\in H_X\times H_Y}\sum_{\tau\in S_m\times S_p} \cr
    \times \sum_{l\in S_{j,m}} \sum_{w\in S'_{i,m}}\delta(\tau^{-1}(w),l)\delta(\tau^{-1}(q),q)
    \delta(\tau^{-1}\beta_2\sigma_2\beta_1^{-1})\,
    \delta(\tau\gamma_2\sigma_1\gamma_1^{-1})\cr
  ={n_{ii}^X(\sigma_1)n_{jj}^X(\sigma_2)\over |H_X\times H_Y| |H'_X\times H'_Y|}
    \sum_{R'} \sqrt{c_{RR'}c_{TR'}}{{\rm hooks}_{R'}\over \sqrt{{\rm hooks}_R {\rm hooks}_T}}\delta_{rw}
    \sum_{\gamma_1,\gamma_2\in H_X'\times H_Y'}\sum_{\beta_1,\beta_2\in H_X\times H_Y} \cr
    \times \sum_{l\in S_{j,m}} \sum_{w\in S'_{i,m}}\delta(\gamma_2\sigma_1\gamma_1^{-1}(w),l)
    \delta(\gamma_2\sigma_1\gamma_1^{-1}(q),q)\,
    \delta(\beta_2\sigma_2\beta_1^{-1}\gamma_2\sigma_1\gamma_1^{-1})\, .\cr
\eea
Consider 
\bea
    \sum_{l\in S_{j,m}} \sum_{w\in S'_{i,m}}\delta(\gamma_2\sigma_1\gamma_1^{-1}(w),l)=n_{ij}^{Y+}(\sigma_1)
\eea
where $n_{ij}^{Y+}(\sigma_2)=n_{ij}^{Y+}(\sigma_1)$ is the number of strings going from $i$ to $j$ in the Gauss graph associated to the $Y$s.
The fact that this $n_{ij}^{Y+}(\sigma_2)$ appears suggests that the strings stretching between $i$ and $j$ in the Gauss graph associated
to the $Y$s are participating, even though it is the $X$ Gauss graph that undergoes the transition.
Thus
\bea
A  ={n_{ii}^X(\sigma_1)n_{jj}^X(\sigma_2)\over |H_X\times H_Y| |H'_X\times H'_Y|}
    \sum_{R'} \sqrt{c_{RR'}c_{TR'}}{{\rm hooks}_{R'}\over \sqrt{{\rm hooks}_R {\rm hooks}_T}}\delta_{rw}
    \sum_{\gamma_1,\gamma_2\in H_X'\times H_Y'}\sum_{\beta_1,\beta_2\in H_X\times H_Y} \cr
    n_{ij}^{Y+}(\sigma_2)
    \delta(\gamma_2\sigma_1\gamma_1^{-1}(q),q)\,
    \delta(\beta_2\sigma_2\beta_1^{-1}\gamma_2\sigma_1\gamma_1^{-1})\cr
\eea
Next, consider the role of $\delta(\gamma_2\sigma_1\gamma_1^{-1}(q),q)$. 
This tells us that a single string, which loops back to the same brane, is plucked from brane $i$ (or $j$) and reattached to brane $j$ (or $i$).
This follows because the string which is plucked has startpoint $q$ and end point $\gamma_2\sigma_1\gamma_1^{-1}(q)$.
So the delta function is setting the start point equal to the end point. 
Another way to say it is that states with different values for the $n_{ij}^Y$ or $n_{ij}^X$ don't mix - and this is why the terms 
in the dilatation operator that mix $X$ and $Y$ commute with terms that mix $X$ and $Z$ and the terms that mix $Y$ and $Z$. 
The role of this delta function is also easy to interpret in terms of the Gauss graph: the two Gauss graphs that mix, $\sigma_1$ and $\sigma_2$, 
are related by peeling a closed loop from node $i$ of $\sigma_1^X$ (or $\sigma_1^Y$) and reattaching it to node $j$ of $\sigma_2^X$
(or $\sigma_2^Y$). 
This implies that, as permutations, $\sigma_1$ and $\sigma_2$ are identical (recall that closed loops are 1 cycles).

If we peel a string from node $i$ of $\sigma_1^X$ and reattach it to node $j$ of $\sigma_2^X$, the factor 
$n_{ii}^X(\sigma_1)n_{jj}^X(\sigma_2)=n_{ii}^X(\sigma_1)(n_{jj}^X(\sigma_1)+1)=(n_{ii}^X(\sigma_2)+1)n_{jj}^X(\sigma_2)$ 
is the number of strings starting and ending at node $i$ before we peel a string off, multiplied by the number of strings starting and ending
at node $j$ after we have attached the string.

Notice that the delta function $\delta(\gamma_2\sigma_1\gamma_1^{-1}(q),q)$ reduces the full sum over $\gamma_1$ and $\gamma_2$ to those
elements of $H_X'\times H_Y'$ that leave $q$ inert. This is a subgroup of $(H_X'\times H_Y')\cap (H_X\times H_Y)$ that we will denote
${\cal G}_{\sigma_1,q}$.
Consequently, the size of this matrix element is
\bea
   n_{ii}^X(\sigma_1)n_{jj}^X(\sigma_2)n_{ij}^{Y+}(\sigma_2)|{\cal G}_{\sigma_1,q}|
\eea
Notice that
\bea
  {|{\cal G}_{\sigma_1,q}|\over N_{\sigma_1}}=n_{ii}^X(\sigma_1)\qquad
  {|{\cal G}_{\sigma_1,q}|\over N_{\sigma_2}}= n_{jj}^X(\sigma_2)\, .
\eea
These two formulas follow because $N_\sigma$ counts the number of symmetries of the Gauss graph, while $|{\cal G}_{\sigma,q}|$ counts the number 
of symmetries that don't include permutations of the closed loop corresponding to $q$. 
Thus, we finally see that the normalized matrix element is nothing but
\bea
  \sqrt{c_{RR'}c_{TR'}}{{\rm hooks}_{R'}\over \sqrt{{\rm hooks}_R {\rm hooks}_T}}\delta_{rw}
  \sqrt{n_{ii}^X(\sigma_1)n_{jj}^X(\sigma_2)}n_{ij}^{Y+}(\sigma_2)\, .
\eea

The evaluation of the fourth term is practically identical and will not be discussed.

\subsection{Second Term}

Now consider the second term
\bea
B={|H_X\times H_Y||H'_X\times H'_Y|\over m!p!}\sum_{jk}\sum_{s\vdash m}\sum_{t\vdash p}\sum_{\vec{\mu}\vec{\nu}}
                                             \sum_{lm}\sum_{x\vdash m}\sum_{y\vdash p}\sum_{\vec{\alpha}\vec{\beta}}                                             
                                             \Gamma^{(s,t)}(\sigma_1)_{jk}\Gamma^{(x,y)}(\sigma_2)_{lm}\cr
\times B^{(s,t)\to 1_{H_X\times H_Y}}_{j\vec{\mu}}B^{(s,t)\to 1_{H_X\times H_Y}}_{k\vec{\nu}}
       B^{(x,y)\to 1_{H'_X\times H'_Y}}_{l\vec{\alpha}}B^{(x,y)\to 1_{H'_X\times H'_Y}}_{m\vec{\beta}}\cr
\sum_{R'} c_{RR'} {d_T mp\over (n+m+p)d_{R'}}
    \sqrt{f_T {\rm hooks}_T\over f_R {\rm hooks}_R}\delta_{rw}
  \langle\vec{p},t,\nu_1;a   |\vec{p}',y,\alpha_1;b\rangle
  \langle\vec{p}',y,\beta_1;b|E^{(1)}_{ji}  |\vec{p},t,\mu_1;a\rangle\cr
  \qquad\times
  \langle\vec{m},s,\nu_2;c   |E^{(p+1)}_{ij}|\vec{m}',x,\alpha_2;d\rangle
  \langle\vec{m}',x,\beta_2;d|\vec{m},s,\mu_2;c\rangle\cr
={1\over |H_X\times H_Y|^2 |H'_X\times H'_Y|^2 m!p!}\cr
\sum_{R'} c_{RR'} {d_T mp\over (n+m+p)d_{R'}}
    \sqrt{f_T {\rm hooks}_T\over f_R {\rm hooks}_R}\delta_{rw}
\sum_{\tau\in S_m\times S_p} 
   \sum_{\phi\in S_m\times S_p}
   \sum_{\gamma_1,\gamma_2\in H_X'\times H_Y'}
   \sum_{\beta_1,\beta_2\in H_X\times H_Y}\cr
\times \delta_{\vec{m}\vec{m}'}\delta_{\vec{p}\vec{p}'}
\langle v,\vec{p}',\vec{m}'|\tau^{-1} \, E^{(1)}_{ji} \,\phi\beta_2\sigma_2\beta_1^{-1} |v,\vec{p},\vec{m}\rangle \cr
\times \langle v,\vec{p},\vec{m}|\phi^{-1} \, E^{(p+1)}_{ij}\,\tau\gamma_2\sigma_1\gamma_1^{-1}|v,\vec{p}',\vec{m}'\rangle\, .\cr
\eea
For the above result to be nonzero it is clear that we need
\bea
   \vec{p}=\vec{p}'\qquad \vec{m}-\hat{i}+\hat{j}=\vec{m}'
\eea
as well as
\bea
   \vec{p}=\vec{p}'\qquad \vec{m}=\vec{m}'\, .
\eea
Consequently, this term is only non-zero when $i=j$. In this case
\bea
B={1\over |H_X\times H_Y|^4 m!p!}\cr
\sum_{R'} c_{RR'} {d_T mp\over (n+m+p)d_{R'}}
    \sqrt{f_T {\rm hooks}_T\over f_R {\rm hooks}_R}\delta_{rw}
\sum_{\tau\in S_m\times S_p} 
   \sum_{\phi\in S_m\times S_p}
   \sum_{\gamma_1,\gamma_2,\gamma\in H_X\times H_Y}
   \sum_{\beta_1,\beta_2,\beta\in H_X\times H_Y}\cr
\times \sum_{k\in S_{j,m}}\delta(\tau^{-1}(1),k)
\delta(\tau^{-1} \phi\beta_2\sigma_2\beta_1^{-1}\beta)\cr
\times \sum_{l\in S_{i,m}}\delta (\phi^{-1}(p+1),l)
\delta (\phi^{-1}\tau\gamma_2\sigma_1\gamma_1^{-1}\gamma)\cr
={1\over |H_X\times H_Y|^2 m!p!}\cr
\sum_{R'} c_{RR'} {d_T mp\over (n+m+p)d_{R'}}
    \sqrt{f_T {\rm hooks}_T\over f_R {\rm hooks}_R}\delta_{rw}
\sum_{\tau\in S_m\times S_p} 
   \sum_{\phi\in S_m\times S_p}
   \sum_{\gamma_1,\gamma_2\in H_X\times H_Y}
   \sum_{\beta_1,\beta_2\in H_X\times H_Y}\cr
\times \sum_{k\in S_{j,m}}\delta(\tau^{-1}(1),k)
\delta(\tau^{-1} \phi\beta_2\sigma_2\beta_1^{-1})\cr
\times \sum_{l\in S_{i,m}}\delta (\phi^{-1}(p+1),l)
\delta (\phi^{-1}\tau\gamma_2\sigma_1\gamma_1^{-1})\cr
={1\over |H_X\times H_Y|^2 m!p!}\cr
\sum_{R'} c_{RR'} {d_T \over (n+m+p)d_{R'}}
    \sqrt{f_T {\rm hooks}_T\over f_R {\rm hooks}_R}\delta_{rw}
\sum_{\tau\in S_m\times S_p} 
   \sum_{\phi\in S_m\times S_p}
   \sum_{\gamma_1,\gamma_2\in H_X\times H_Y}
   \sum_{\beta_1,\beta_2\in H_X\times H_Y}\cr
\times 
\sum_{k\in S_{j,m}}\sum_{r=1}^m\delta(\tau^{-1}(m),k)
\sum_{l\in S_{i,m}}\sum_{s=p+1}^{p+m}\delta (\phi^{-1}(s),l)
\delta(\tau^{-1} \phi\beta_2\sigma_2\beta_1^{-1})
\delta (\phi^{-1}\tau\gamma_2\sigma_1\gamma_1^{-1})\cr
={1\over |H_X\times H_Y|^2}\cr
\sum_{R'} c_{RR'} {d_T p\over (n+m+p)d_{R'}}
    \sqrt{f_T {\rm hooks}_T\over f_R {\rm hooks}_R}\delta_{rw}
\sum_{\tau\in S_m\times S_p} 
   \sum_{\gamma_1,\gamma_2\in H_X\times H_Y}
   \sum_{\beta_1,\beta_2\in H_X\times H_Y}\cr
\times n^{+,X}_{i}n^{+,Y}_{i}
\delta(\tau^{-1}\beta_2\sigma_2\beta_1^{-1})
\delta (\tau\gamma_2\sigma_1\gamma_1^{-1})\cr
={1\over |H_X\times H_Y|^2}\cr
\sum_{R'} c_{RR'} {d_T \over (n+m+p)d_{R'}}
    \sqrt{f_T {\rm hooks}_T\over f_R {\rm hooks}_R}\delta_{rw}
   \sum_{\gamma_1,\gamma_2\in H_X\times H_Y}
   \sum_{\beta_1,\beta_2\in H_X\times H_Y}\cr
\times n^{+,X}_{i}n^{+,Y}_{i}
\delta (\beta_2\sigma_2\beta_1^{-1}\gamma_2\sigma_1\gamma_1^{-1})\cr
=\sum_{R'} c_{RR'} {d_T \over (n+m+p)d_{R'}}
    \sqrt{f_T {\rm hooks}_T\over f_R {\rm hooks}_R}\delta_{rw}
    \sum_{\beta_1,\beta_2\in H_X\times H_Y} n^{+,X}_{ii}n^{+,Y}_{ii}
\delta (\beta_2\sigma_2\beta_1^{-1}\sigma_1)\cr
\eea
In summary, and perhaps writing it a bit more clearly, we have
\bea
B= {|H_X\times H_Y||H'_X\times H'_Y|\over m!p!}\sum_{jk}\sum_{s\vdash m}\sum_{t\vdash p}\sum_{\vec{\mu}\vec{\nu}}
  \sum_{lm}\sum_{x\vdash m}\sum_{y\vdash p}\sum_{\vec{\alpha}\vec{\beta}}\Gamma^{(s,t)}(\sigma_1)_{jk}\Gamma^{(x,y)}(\sigma_2)_{lm}\cr
  \times B^{(s,t)\to 1_{H_X\times H_Y}}_{j\vec{\mu}}B^{(s,t)\to 1_{H_X\times H_Y}}_{k\vec{\nu}}
  B^{(x,y)\to 1_{H'_X\times H'_Y}}_{l\vec{\alpha}}B^{(x,y)\to 1_{H'_X\times H'_Y}}_{m\vec{\beta}}\cr
  \sum_{R'} c_{RR'} {d_T mp\over (n+m+p)d_{R'}}
  \sqrt{f_T {\rm hooks}_T\over f_R {\rm hooks}_R}\delta_{rw}
  \langle\vec{p},t,\nu_1;a   |\vec{p}',y,\alpha_1;b\rangle
  \langle\vec{p}',y,\beta_1;b|E^{(1)}_{ji}  |\vec{p},t,\mu_1;a\rangle\cr
  \qquad\times
  \langle\vec{m},s,\nu_2;c   |E^{(p+1)}_{ij}|\vec{m}',x,\alpha_2;d\rangle
  \langle\vec{m}',x,\beta_2;d|\vec{m},s,\mu_2;c\rangle\cr
 =\delta_{\vec{p}\vec{p}'}\delta_{\vec{m}\vec{m}'}\sum_{R_i'} c_{RR_i'} {d_T \over (n+m+p)d_{R'}}
  \sqrt{f_T {\rm hooks}_T\over f_R {\rm hooks}_R}\delta_{rw}
  \sum_{\beta_1,\beta_2\in H_X\times H_Y} n^{+,X}_{i}n^{+,Y}_{i}
  \delta (\beta_2\sigma_2\beta_1^{-1}\sigma_1)\cr
\eea
Notice that this term is already diagonal in the Gauss graph basis.
Notation: $n^{+,X}_{i}$ is the number of strings ending on node $i$ of the $X$ Gauss graph;
$n^{X}_{ii}$ is the number of strings starting on and then looping back to end on node $i$ of the $X$ Gauss graph.

The evaluation of the third term is practically identical and will not be discussed.

\subsection{Final Answer}

In this section we will summarize the action of the term in the dilatation operator that mixes $X$s and $Y$s on the Gauss operators.
\begin{figure}[ht]%
\begin{center}
\includegraphics[width=15cm]{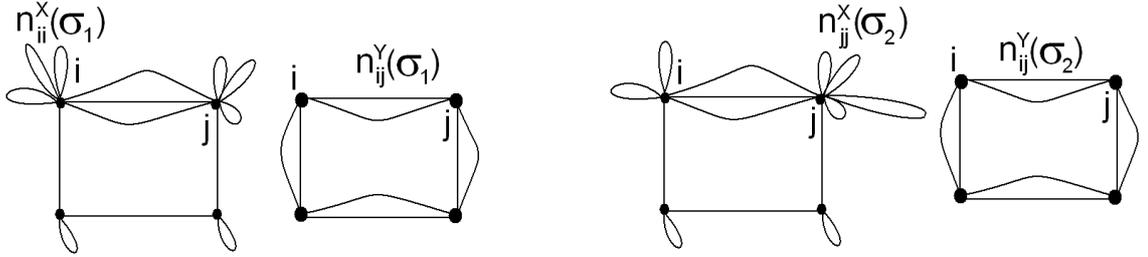}%
\caption{The Gauss graph on the left is described by $\sigma_1$, while the Gauss graph on the right is described by $\sigma_2$.
         To make a transition between the two pairs of Gauss graphs shown, we pluck a string from node $i$ of the $X$ graph 
         on the left and attach it to node $j$ of the $X$ graph on the right. The numbers which participate are (i)
         the number of strings $n_{ij}^Y$ stretching between nodes $i$ and $j$ of the $Y$ graph, (ii) the number of strings attached to node
         $i$ of the $X$ graph before a string is removed $n_{ii}^X(\sigma_1)=n_{ii}^X(\sigma_2)+1$ and (iii) the number of strings attached
         to the node $j$ of the $X$ graph after a string is attached $n_{ii}^X(\sigma_1)+1=n_{ii}^X(\sigma_2)$.}
\label{ggraph}%
\end{center}
\end{figure}

Here is the final answer for matrix elements of $D$ taken with {\it normalized operators}. The diagonal terms are
\bea
\langle O^\dagger_{R,r}(\sigma )D_{XY}O_{R,r}(\sigma)\rangle = 
                2\sum_{i=1}^p {c_{RR_i'}\over l_{i}}\left(n(\sigma)_{i}^{+\, X}n(\sigma)_{i}^{+\, Y}
                -n(\sigma)_{ii}^{+\, X}n(\sigma)_{ii}^{+\, Y}\right)
\label{ddelem1}
\eea
Now, consider an off diagonal term. 
One possible non-zero matrix element corresponds to the case that the $X$ Gauss graph changes,
by detaching a loop from node $i$ of the $\sigma_1^X$ Gauss graph and reattaching it to node $j$.
The matrix element describing this process is (recall that we only ever get a non-zero matrix element if
$n_{ij}^Y(\sigma_1)=n_{ij}^Y(\sigma_2)$ and $n_{ij}^X(\sigma_1)=n_{ij}^X(\sigma_2)$)
\bea
  \langle O^\dagger_{R,r}(\sigma_1 )D_{XY}O_{R,r}(\sigma_2 )\rangle \,\, =
  \,\, -\sqrt{c_{RR'}c_{TR'}\over l_{i}l_{j}}n_{ij}^Y(\sigma_1) \sqrt{n_{ii}^X(\sigma_1) (n_{jj}^{X}(\sigma_1)+1)}
\label{ddelem2}
\eea
Another non-zero matrix element is obtained when the $Y$ Gauss graph changes,
by detaching a loop from node $i$ of the $\sigma_1^Y$ Gauss graph and reattaching it to node $j$.
The matrix element describing this process is 
\bea
  \langle O^\dagger_{R,r}(\sigma_1 )D_{XY}O_{R,r}(\sigma_2 )\rangle \,\, =
  \,\,-\sqrt{c_{RR'}c_{TR'}\over l_{i}l_{j}}n_{ij}^X\sqrt{n_{ii}^{Y}(\sigma_1) (n_{jj}^{Y}(\sigma_1)+1)}
\label{ddelem3}
\eea
This gives a complete description of the action of the term in the dilatation operator that mixes $X$s and $Y$s on the Gauss operators.

\section{Diagonalization}

\begin{figure}[ht]%
\begin{center}
\includegraphics[width=15cm]{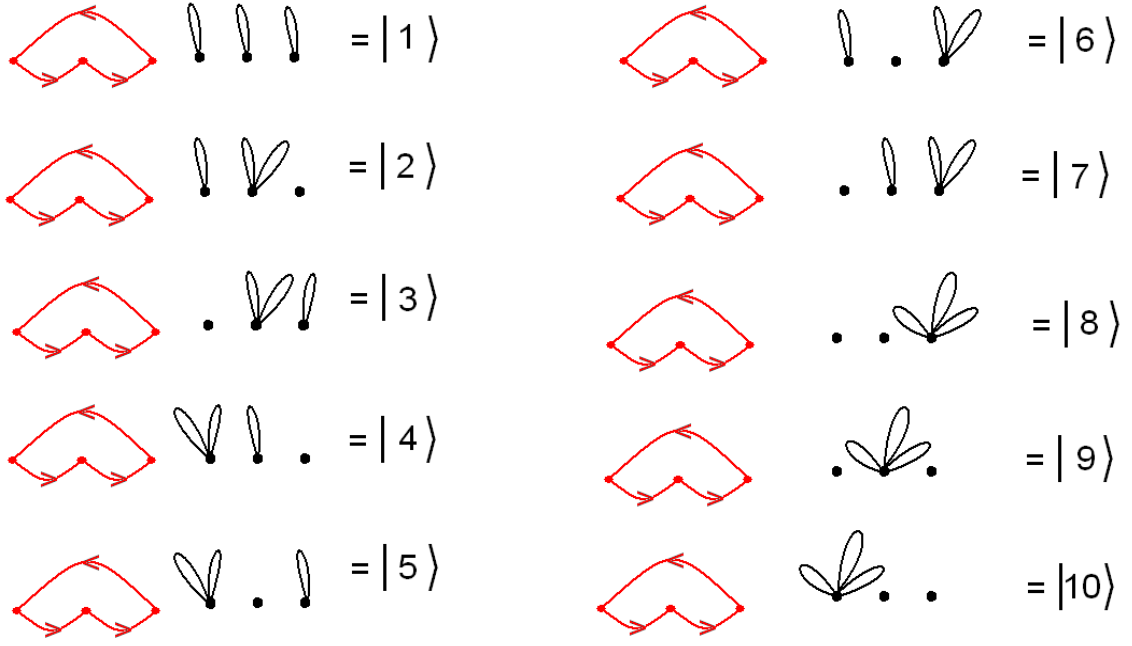}%
\caption{The 10 states that appear in our first example are defined in the figure above.}
\label{nice}%
\end{center}
\end{figure}

To understand the structure of the diagonalization problem, lets start off with a warm up problem.
This will also be an example of the use of the formulas (\ref{ddelem1}), (\ref{ddelem2}) and (\ref{ddelem3}), which will allow the
reader to test her understanding of our result.
Consider the Gauss graphs shown in figure \ref{nice}.
Using the formulas from the previous section, there is a transition between $|1\rangle$ and $|2\rangle$.
To understand how we have labeled the dots, we must detach a loop from black node 3 of $|1\rangle$ and attach it to black node 2 of $|2\rangle$.
Denote the Gauss graph correspodning to $|1\rangle$ by $\sigma_1$ and the Gauss graph of $|2\rangle$ by $\sigma_2$.
We have $n_{23}^X(\sigma_1)=1$ (read from the red Gauss graph), $n_{22}^{Y}(\sigma_1)+1=2$ read from $|1\rangle$ and 
$n_{33}^{Y}(\sigma_1)=1$ read from $|1\rangle$.
Thus, in total the matrix element is
\bea
   -\sqrt{(N+l_2)(N+l_3)\over l_2 l_3}\sqrt{2}
\eea
As a second example, the matrix element for the transition between $|2\rangle$ and $|3\rangle$ is
\bea
   -\sqrt{(N+l_1)(N+l_3)\over l_1 l_3}
\eea
For the 10 states shown, we have the off diagonal piece of the dilatation operator given by
\bea
  -\sqrt{(N+l_1)(N+l_2)\over l_1 l_2}M_{12}-\sqrt{(N+l_2)(N+l_3)\over l_2 l_3}M_{23}-\sqrt{(N+l_1)(N+l_3)\over l_1 l_3}M_{13}
\eea
where
\bea
   M_{12}=\left[
\begin{array}{cccccccccc}
0 &0 &-\sqrt{2} &0 &-\sqrt{2} &0 &0 &0 &0 &0\cr
0 &0 &0 &-2 &0 &0 &0 &0 &-\sqrt{3} &0\cr
-\sqrt{2} &0 &0 &0 &0 &0 &0 &0 &0 &0\cr
0 &-2 &0 &0 &0 &0 &0 &0 &0 &-\sqrt{3}\cr
-\sqrt{2} &0 &0 &0 &0 &0 &0 &0 &0 &0\cr
0 &0 &0 &0 &0 &0 &-1 &0 &0 &0\cr
0 &0 &0 &0 &0 &-1 &0 &0 &0 &0\cr
0 &0 &0 &0 &0 &0 &0 &0 &0 &0\cr
0 &-\sqrt{3} &0 &0 &0 &0 &0 &0 &0 &0\cr
0 &0 &0 &-\sqrt{3} &0 &0 &0 &0 &0 &0
\end{array}
\right]
\eea
\bea
   M_{23}=\left[
\begin{array}{cccccccccc}
0 &-\sqrt{2} &0 &0 &0 &-\sqrt{2} &0 &0 &0 &0\cr
-\sqrt{2} &0 &0 &0 &0 &0 &0 &0 &0 &0\cr
0 &0 &0 &0 &0 &0 &-2 &0 &-\sqrt{3} &0\cr
0 &0 &0 &0 &-1 &0 &0 &0 &0 &0\cr
0 &0 &0 &-1 &0 &0 &0 &0 &0 &0\cr
-\sqrt{2} &0 &0 &0 &0 &0 &0 &0 &0 &0\cr
0 &0 &-2 &0 &0 &0 &0 &-\sqrt{3} &0 &0\cr
0 &0 &0 &0 &0 &0 &-\sqrt{3} &0 &0 &0\cr
0 &0 &-\sqrt{3} &0 &0 &0 &0 &0 &0 &0\cr
0 &0 &0 &0 &0 &0 &0 &0 &0 &0
\end{array}
\right]
\eea
\bea
   M_{13}=\left[
\begin{array}{cccccccccc}
0 &0 &0 &-\sqrt{2} &0 &0 &-\sqrt{2} &0 &0 &0\cr
0 &0 &-1 &0 &0 &0 &0 &0 &0 &0\cr
0 &-1 &0 &0 &0 &0 &0 &0 &0 &0\cr
-\sqrt{2} &0 &0 &0 &0 &0 &0 &0 &0 &0\cr
0 &0 &0 &0 &0 &-2 &0 &0 &0 &-\sqrt{3}\cr
0 &0 &0 &0 &-2 &0 &0 &-\sqrt{3} &0 &0\cr
-\sqrt{2} &0 &0 &0 &0 &0 &0 &0 &0 &0\cr
0 &0 &0 &0 &0 &-\sqrt{3} &0 &0 &0 &0\cr
0 &0 &0 &0 &0 &0 &0 &0 &0 &0\cr
0 &0 &0 &0 &-\sqrt{3} &0 &0 &0 &0 &0
\end{array}
\right]
\eea
and we have the on diagonal piece of the dilatation operator given by
\bea
   {(N+l_1)\over l_1}M_{11}+{(N+l_2)\over l_2}M_{22}+{(N+l_3)\over l_3}M_{13}
\eea
where
\bea
   M_{11}=\left[
\begin{array}{cccccccccc}
2 &0 &0 &0 &0 &0 &0 &0 &0 &0\cr
0 &2 &0 &0 &0 &0 &0 &0 &0 &0\cr
0 &0 &0 &0 &0 &0 &0 &0 &0 &0\cr
0 &0 &0 &4 &0 &0 &0 &0 &0 &0\cr
0 &0 &0 &0 &4 &0 &0 &0 &0 &0\cr
0 &0 &0 &0 &0 &2 &0 &0 &0 &0\cr
0 &0 &0 &0 &0 &0 &0 &0 &0 &0\cr
0 &0 &0 &0 &0 &0 &0 &0 &0 &0\cr
0 &0 &0 &0 &0 &0 &0 &0 &0 &0\cr
0 &0 &0 &0 &0 &0 &0 &0 &0 &6
\end{array}
\right]
\eea
\bea
   M_{22}=\left[
\begin{array}{cccccccccc}
2 &0 &0 &0 &0 &0 &0 &0 &0 &0\cr
0 &4 &0 &0 &0 &0 &0 &0 &0 &0\cr
0 &0 &4 &0 &0 &0 &0 &0 &0 &0\cr
0 &0 &0 &2 &0 &0 &0 &0 &0 &0\cr
0 &0 &0 &0 &0 &0 &0 &0 &0 &0\cr
0 &0 &0 &0 &0 &0 &0 &0 &0 &0\cr
0 &0 &0 &0 &0 &0 &2 &0 &0 &0\cr
0 &0 &0 &0 &0 &0 &0 &0 &0 &0\cr
0 &0 &0 &0 &0 &0 &0 &0 &6 &0\cr
0 &0 &0 &0 &0 &0 &0 &0 &0 &0
\end{array}
\right]
\eea
\bea
   M_{33}=\left[
\begin{array}{cccccccccc}
2 &0 &0 &0 &0 &0 &0 &0 &0 &0\cr
0 &0 &0 &0 &0 &0 &0 &0 &0 &0\cr
0 &0 &2 &0 &0 &0 &0 &0 &0 &0\cr
0 &0 &0 &0 &0 &0 &0 &0 &0 &0\cr
0 &0 &0 &0 &2 &0 &0 &0 &0 &0\cr
0 &0 &0 &0 &0 &4 &0 &0 &0 &0\cr
0 &0 &0 &0 &0 &0 &4 &0 &0 &0\cr
0 &0 &0 &0 &0 &0 &0 &6 &0 &0\cr
0 &0 &0 &0 &0 &0 &0 &0 &0 &0\cr
0 &0 &0 &0 &0 &0 &0 &0 &0 &0
\end{array}
\right]
\eea
To get some insight into the structure of these matrices, note that the matrix
\bea
   M=M_{11}+M_{22}+M_{33}+M_{12}+M_{23}+M_{13}
\eea
has eigenvalues $0,3,3,6,6,6,9,9,9,9$. 
The even spacing and the degeneracy of the eigenvalues matches the weights of the symmetric representation $\yng(3)$ of $SU(3)$.
This strongly suggests that, we can understand the off diagonal pieces of the dilatation operator as raising/lowering operators
of some $SU(k)$ representations, with $k\le g$. 
Recall that $g$ is the number of rows in our restricted Schur polynomials.
This guess turns out to be correct as we now explain.

First, we need to define a bijection between the Gauss graphs that mix and the states of a particular unitary group representation.
Lets start by considering a situation for which the $Y$ Gauss graph is fixed and we have transitions between different $X$ Gauss graphs.
We can only have transitions of closed loops between nodes $i$ and $j$ in the $X$ Gauss graph if $n_{ij}^Y(\sigma)\ne 0$.
Denote the number of connected components of $\sigma^Y$ by $C$.
Each connected component is a set of directed line segments running between nodes.
Let $c$ denote the number of connected components that have more than a single node.
Let the number of nodes in each of these connected components be $n_i$, $i=1,...,c$.
The irreducible representation that organizes the $\sigma^X$ graphs is an irreducible representation of the group
\bea
   SU(n_1)\times SU(n_2)\times \cdots \times SU(n_c)
\eea
Focus on one of the connected components, say the $j^{\rm th}$ connected component. 
Assume that there are a total of $\tilde{n}$ closed loops attached to nodes of $\sigma^X$ that belong to this connected component.
The irreducible representation of the $SU(n_j)$ factor in the above group that plays a role is labeled by a Young diagram that has a
single row containing $\tilde{n}$ boxes.
We now want to give the map between different $X$ Gauss graphs and states of this representation.
Number the nodes in the $j^{\rm th}$ connected component from 1 up to $n_j$.
Consider an $X$ Gauss graph that has $n_{11}$ strings attached to node 1, $n_{22}$ to node 2, and so on up to $n_{n_jn_j}$ attached to node $n_j$.
The Gelfand-Tsetlin pattern for this state
$$
{\footnotesize
|M\rangle =\left[ 
\begin {array}{ccccccc} 
m_{1,n_j} &              & m_{2,n_j}  & \dots       & m_{n_j-1,n_j} &                 & m_{n_jn_j}\\\noalign{\medskip}
          & m_{1,n_j-1}  &            & m_{2,n_j-1} & \dots         & m_{n_j-1,n_j-1} &       \\\noalign{\medskip}
          &              & \dots      &\dots        &  \dots        &                 &       \\\noalign{\medskip}
          &              &m_{1,2}     &             &m_{2,2}        &                 &       \\\noalign{\medskip}
          &              &            &m_{1,1}      &               &                 &
\end {array} \right] }
$$
has $m_{p,q}=0$ for $p>1$ and
\bea
   m_{1,q}=\sum_{i=1}^q n_{ii}
\eea
This completes our discussion of how the Gauss graphs are organized, for a fixed $\sigma^Y$.
To complete the discussion note that there is a completely parallel argument with the roles of $\sigma^X$ and $\sigma^Y$ switched.

\begin{figure}[ht]%
\begin{center}
\includegraphics[width=15cm]{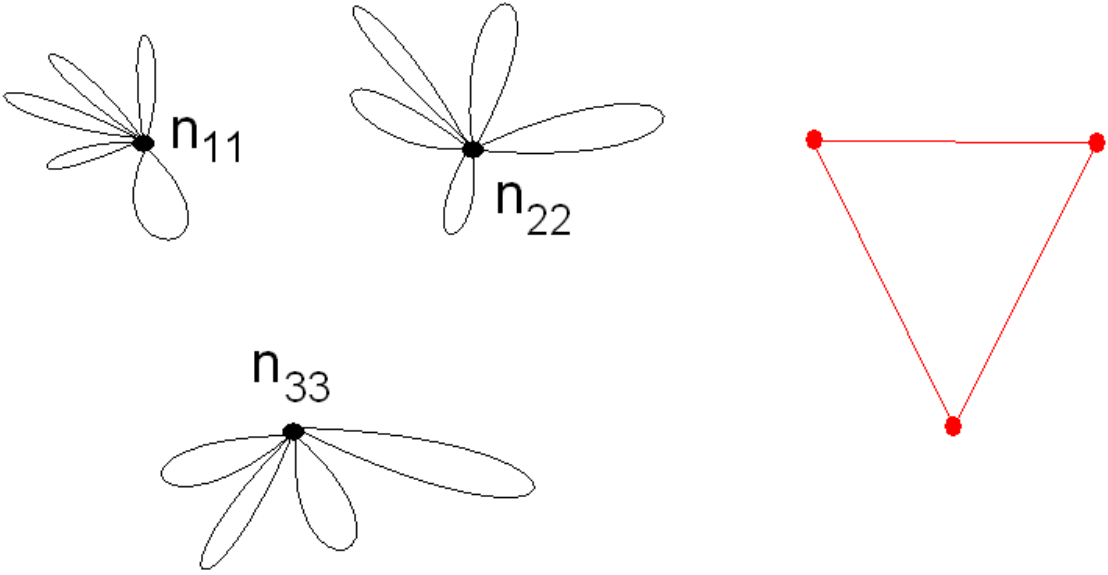}%
\caption{The Gauss graph is shown in black. Closed loops can detach from a node and reattach to another node.}
\label{GT}%
\end{center}
\end{figure}

As a concrete example, the Gauss graph in figure \ref{GT} corresponds to the Gelfand-Tsetlin pattern
$$
{\footnotesize
|M\rangle =\left[ 
\begin {array}{ccccc} 
n_{11}+n_{22}+n_{33} &                & 0        &   & 0 \\\noalign{\medskip}
                     & n_{11}+n_{22}  &          & 0 &   \\\noalign{\medskip}
                     &                & n_{11}   &   &   
\end {array} \right] }
$$
This map between Gelfand-Tsetlin patterns and operators labeled by Gauss graphs turns out to be useful because we know the matrix
elements of the Lie algebra elements in the Gelfand-Tsetlin basis. 
For example, let us consider the lowering operator $E_{i,i+1}$.
This will shift $n_{ii}\to n_{ii}-1$ and $n_{i+1,i+1}\to n_{i+1,i+1}+1$.
The net effect of these shifts in the Gelfand-Tsetlin pattern is to replace $m_{i,k}\to m_{i,k}-1$; we will denote this pattern by $M^-_i$.
The Gauss graph corresponding to $M^-_i$ is obtained from the Gauss graph corresponding to $M$ by peeling a closed loop from
node $i$ and reattaching it to node $i+1$.
We have already studied the matrix element of the dilatation operator that mixes these two Gauss graphs and have found 
\bea
  -\sqrt{c_{RR'}c_{TR'}\over l_{i}l_{i+1}}n_{i,i+1}^Y(\sigma)\sqrt{n_{ii}^{X}(\sigma) (n_{i+1,i+1}^{X}(\sigma)+1)}
  \label{delem}
\eea
where $\sigma$ describes the state with Gelfand-Tsetlin pattern $M$.
According to \cite{BR} the matrix element for the lowering operator, written in terms of the entries of the Gelfand-Tsetlin pattern, is
\bea
   \langle M^-_i|E_{i,i+1}|M\rangle =
\sqrt{-
{\prod_{k'=1}^{l+1}(m_{k',l+1}-m_{k,l}+k-k'+1)\prod_{k'=1}^{l-1}(m_{k',l-1}-m_{k,l}+k-k')\over 
\prod_{k'=1,k'\ne k}^{l}(m_{k',l+1}-m_{k,l}+k-k'+1)(m_{k',l+1}-m_{k,l}+k-k')}}\cr
\label{rlelem}
\eea

Plugging in the patterns for the two Gauss graphs that mix, it is straight forward to see that (\ref{rlelem}) evaluates to
\bea
  \sqrt{n_{ii}^{X}(\sigma) (n_{i+1,i+1}^{X}(\sigma)+1)}
\eea
Comparing to (\ref{delem}) we see that the off diagonal term of the dilatation operator that we are considering is in fact
\bea
  -\sqrt{c_{RR'}c_{TR'}\over l_{i}l_{i+1}}n_{i,i+1}^Y(\sigma) E_{i,i+1}
\eea

\begin{figure}[ht]%
\begin{center}
\includegraphics[width=15cm]{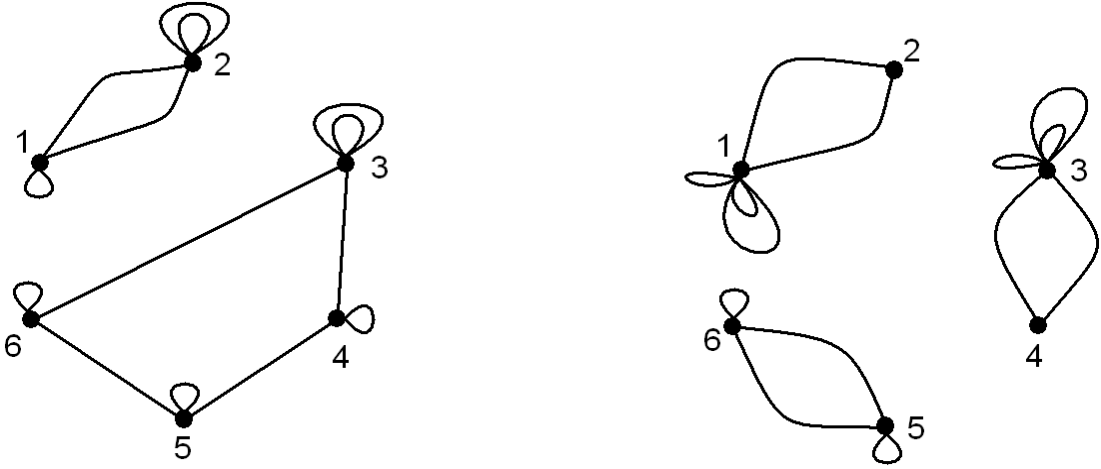}%
\caption{The graph on the left is $\sigma^X$. The graph on the right is $\sigma^Y$. Each node label in the above diagrams corresponds to
         a row number of Young diagram $R$ in the restricted Schur polynomial $\chi_{R,(t,s,r)\vec{\mu}\vec{\nu}}$. The Gauss graphs shown
         correspond to an $R$ with 6 long rows.}
\label{bigGT}%
\end{center}
\end{figure}

We will state the result for the general case, for which loops move on both the $X$ and $Y$ Gauss graphs, using an example for illustration.
The Gauss graph relevant for this example is show in figure \ref{bigGT}.

Note that $\sigma^X$ has two connected components, one which has 2 nodes and one which has 4 nodes. 
Consequently the group relevant for the organization of the $Y$s is $SU(2)\times SU(4)$.
Counting closed loops on the nodes in $\sigma^Y$ grouped by the connected components of $\sigma^X$
we find that the representation of $SU(2)$ we need is $\yng(2)$ while the representation of $SU(4)$ we need is $\yng(5)$. 
Also, $\sigma^Y$ has three connected components, each of which has 2 nodes. 
Consequently the group relevant for the organization of the $Y$s is $SU(2)\times SU(2)\times SU(2)$.
Counting closed loops on the nodes in $\sigma^X$ grouped by the connected components of $\sigma^Y$
we find that the three representations for the three different $SU(2)$ groups we have are $\yng(3)$, $\yng(3)$ and $\yng(2)$. 
Denoting the groups that appear with a superscript
\bea
  G^{(1)}\times G^{(2)}\times G^{(3)}\times G^{(4)}\times G^{(5)} = SU(2)\times SU(4)\times SU(2)\times SU(2)\times SU(2)
\eea
we can write the off diagonal terms in the dilatation operator as (the superscript on the Lie algebra element tells you which group it belongs to)
\bea
  D_{\rm off\,\, diagonal}=
-\sqrt{(N+l_3)(N+l_4)\over l_3 l_4}(E^{(2)}_{12}+E^{(2)}_{21}) 
-\sqrt{(N+l_4)(N+l_5)\over l_4 l_5}(E^{(2)}_{23}+E^{(2)}_{32})\cr 
-\sqrt{(N+l_5)(N+l_6)\over l_5 l_6}(E^{(2)}_{34}+E^{(2)}_{43}) 
-\sqrt{(N+l_6)(N+l_3)\over l_6 l_3}(E^{(2)}_{14}+E^{(2)}_{41})\cr 
-2\sqrt{(N+l_1)(N+l_2)\over l_1 l_2}(E^{(1)}_{12}+E^{(1)}_{21}) 
-2\sqrt{(N+l_3)(N+l_4)\over l_3 l_4}(E^{(3)}_{12}+E^{(3)}_{21})\cr 
-2\sqrt{(N+l_3)(N+l_4)\over l_3 l_4}(E^{(4)}_{12}+E^{(4)}_{21}) 
-2\sqrt{(N+l_3)(N+l_4)\over l_3 l_4}(E^{(5)}_{12}+E^{(5)}_{21}) 
\eea
The specific representation we should use for each Lie algebra has been spelt out above.

\section{Conclusions and Discussion}\label{conclusions}

In this article we have evaluated certain subleading terms in the action of the dilatation operator in the $SU(3)$ sector.
The operators we have studied have a classical dimension that scales as $N$.
Consequently, even at large $N$, non-planar diagrams need to be summed and the limit we study is quite distinct from the planar limit.
There is by now growing evidence that the dilatation operator in the large $N$ but non-planar limit can be mapped into the Hamiltonian
of a set of decoupled oscillators and hence that this limit of the theory continues to enjoy integrability.
In the $SU(2)$ sector, a new conservation law has been found.
The corrections that we have evaluated spoil this new conservation law and consequently, these terms may be the first indiactions that the 
limit we consider is not integrable.

Our results clearly show that although the new terms do spoil the old conservation law, the system that emerges continues to be integrable.
Indeed, the terms in the Hamiltonian that mix $X$ and $Z$ or $Y$ and $Z$ commute with the terms that mix $X$ and $Y$, so that we simply
need to change basis inside eigenspaces of fixed anomalous dimension.
This change of basis has been reduced to the problem of diagonalizing certain elements in the Lie algebra of a well defined representation
of a definite product of special unitary groups (the specific representation and product can be read off of the Gauss graphs as we explained 
in the last section).
This is a solved problem in group theory.

The term in the dilatation operator that mixes $X$ and $Y$ does not act on the $Z$ labels.
The eigenproblem in the $Z$ label, after moving to the Gauss operator basis, reduces to an oscillator problem\cite{gs}.
The eigenvalues of the term in the dilatation operator that mixes $X$ and $Y$ sets the ground state energy of these oscillators.
Note however that the BPS operators, which correspond to Gauss graphs with loops that start and end on the same node but no
directed segments between nodes, are annihilated by the term in the dilatation operator that mixes $X$ and $Y$, so that these operators remain BPS
even when the corrections we have computed are included. 

Finally, it will be interesting to extend our analysis to the $SU(2|3)$ sector (and beyond) and to see if integrability persists.
We leave these projects for the future.

{\vskip 1.0cm}

\noindent
{\it Acknowledgements:} 
This work is supported by the South African Research Chairs
Initiative of the Department of Science and Technology and the National Research Foundation.
Any opinion, findings and conclusions or recommendations expressed in this material
are those of the authors and therefore the NRF and DST do not accept any liability
with regard thereto.
RdMK thanks Perimeter Institute for hospitality and for providing an excellent research environment.
Research at Perimeter Institute is supported by the Government of Canada
through Industry Canada and by the Province of Ontario through the Ministry of Economic Development \&
Innovation.

\end{document}